\documentclass[twoside,leqno,twocolumn]{article}

\usepackage[letterpaper]{geometry}

\usepackage{ltexpprt}
\usepackage[compress]{cite}
\usepackage{hyperref}

\usepackage{algpseudocode}
\algnewcommand\algorithmiccommon{\textbf{Common:}}
\algnewcommand\Common{\item[\algorithmiccommon]}

\usepackage{subcaption}
\usepackage{graphicx}
\usepackage{tikz}
\usepackage{xcolor}

\definecolor{ridgecolor}{HTML}{56B4E9}
\definecolor{starcolor}{HTML}{F0E442}

\newtheorem{observation}{Observation}
\usepackage{float}
\usepackage{amssymb}

\newcommand{\tetrahedron}{%
  \mathord{\raisebox{-0.1em}{\includegraphics[height=0.7em]{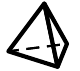}}}%
}

\usepackage{cleveref}
\crefname{line}{Line}{lines}
\Crefname{line}{Line}{Lines}
\crefname{@algorithm}{Alg.}{Algs.}
\Crefname{@algorithm}{Algorithm}{Algorithms}

\crefname{figure}{Fig.}{Figs.}
\Crefname{figure}{Figure}{Figures}

\crefname{table}{Table}{Tables}
\Crefname{table}{Table}{Tables}

\crefname{equation}{Eq.}{Eqs.}
\Crefname{equation}{Equation}{Equations}

\crefname{section}{Section}{Sections}
\Crefname{section}{Section}{Sections}

\makeatletter
\renewcommand{\footnoterule}{%
  \kern-3\p@
  \hrule \@width .4\columnwidth
  \kern 4\p@
}
\makeatother

\providecommand{\submissionID}{XYZ}
\providecommand{\anon}[2]{{#2}}

\usepackage[switch]{lineno}

\begin{document}
\newcommand\relatedversion{}

\title{\Large An advancing-ridge approach for recovering boundary\\$(d-1)$-simplices in $d$-dimensional meshes}
\anon{
    \author{Submission ID \submissionID}
}
{
    \author{Philip Caplan\thanks{Middlebury College, Department of Computer Science.}}    
}

\date{}

\maketitle

\anon{\aftergroup\linenumbers}{}


\begin{abstract} \small\baselineskip=9pt
\anon{Authors are anonymized for double-blind review.}{}
Boundary-conforming four-dimensional meshes are essential for being able to run spacetime numerical simulations about complex, moving three-dimensional geometries.
Specifically, a mesh of pentatopes is needed in which the tetrahedral faces of this mesh conform to the boundary of the domain.
In the three-dimensional setting, a common approach consists of generating a constrained Delaunay tetrahedralization.
This approach typically starts from an unconstrained Delaunay tetrahedralization and recovers the missing boundary segments and triangles by detecting which mesh entities intersect the constraints and then modifies the mesh topology to recover them while enforcing the constrained Delaunay property.
These tetrahedralizations may require additional vertices (Steiner vertices) when a constrained tetrahedralization satisfying all input constraints does not exist.
Implementations of this approach are mature, but it is unclear how it extends to the four-dimensional setting, particularly in how the local mesh operations are scheduled to recover the constraints.
This paper develops a new algorithm for recovering boundary constraints which is simple to implement in any dimension.
The algorithm is primarily an advancing-front approach and uses a constrained cavity operator to incrementally insert constraints into the mesh.
Compared to existing advancing-front approaches, which advance from a front of $(d-1)$-simplices (faces), the proposed approach advances from a front of $(d-2)$-simplices, called ridges.
Steiner vertices can be added to the boundary when the front stalls and several examples in $3d$ demonstrate the ability of this algorithm to recover a complete representation of the input surface.
For the four-dimensional geometries studied here, the algorithm generally recovers at least 99\% of the input tetrahedralization with this advancing ridge procedure.
For some simpler domains, complete conformity with the input tetrahedralization is achieved by adding Steiner vertices, thereby demonstrating the ability to produce boundary-conforming four-dimensional meshes.
The design and efficiency of the underlying cavity operator implementation is also evaluated, showing that 30 million pentatopes can be created in about 1.5 minutes, and 300 million pentatopes in about 15 minutes on a workstation laptop.
\end{abstract}

\section{Introduction}
\begin{figure*}[t]
    \centering
    \begin{subfigure}[T]{0.275\textwidth}
      \begin{tikzpicture}
        \node at (0, 0) {};
        \node at(0, -2) {\includegraphics[width=\textwidth]{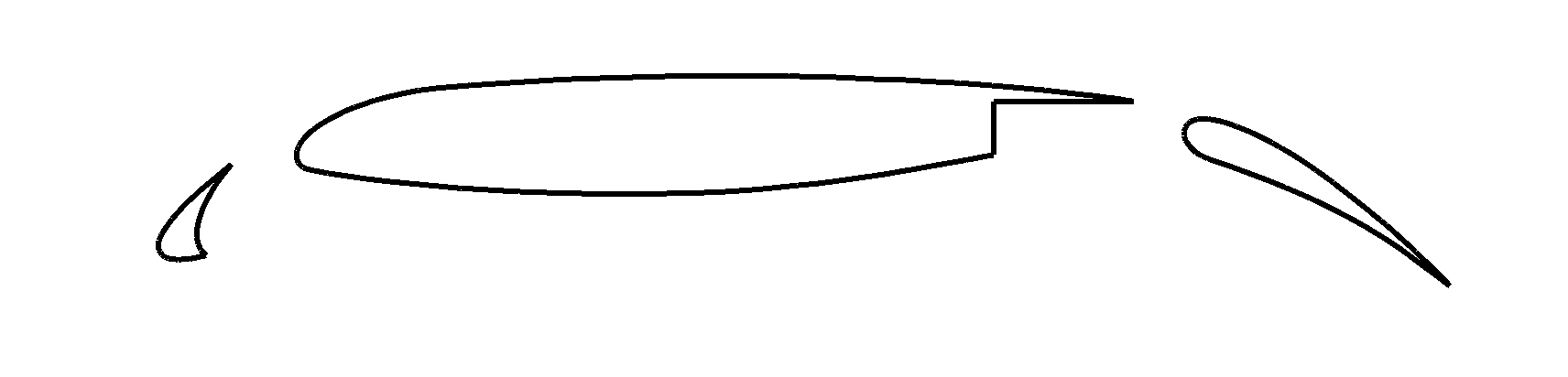}};
        \node at (0, -4) {\includegraphics[width=\textwidth]{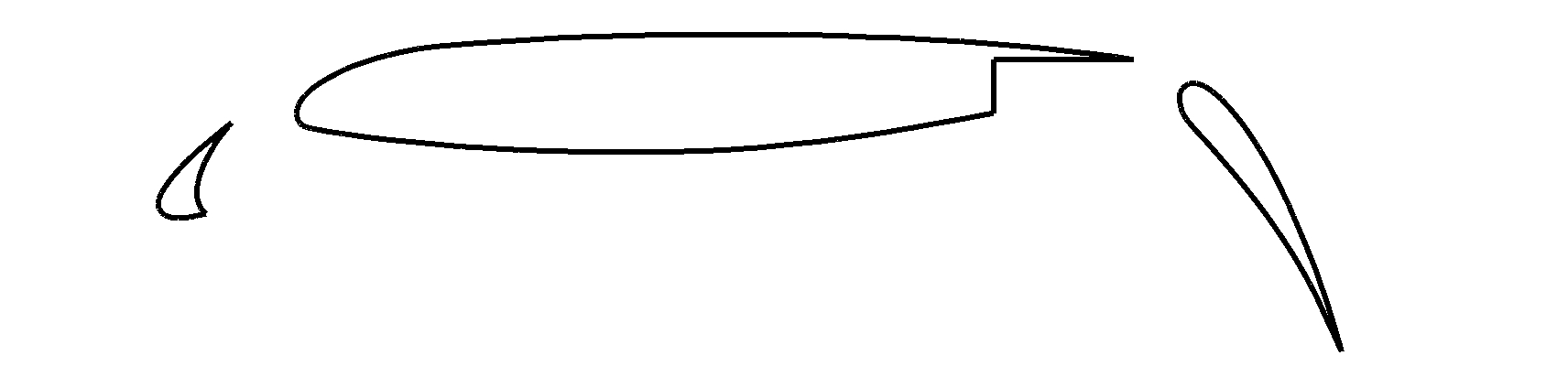}};
      \end{tikzpicture}
    \end{subfigure}
    \begin{subfigure}[T]{0.35\textwidth}
      \includegraphics[width=\textwidth]{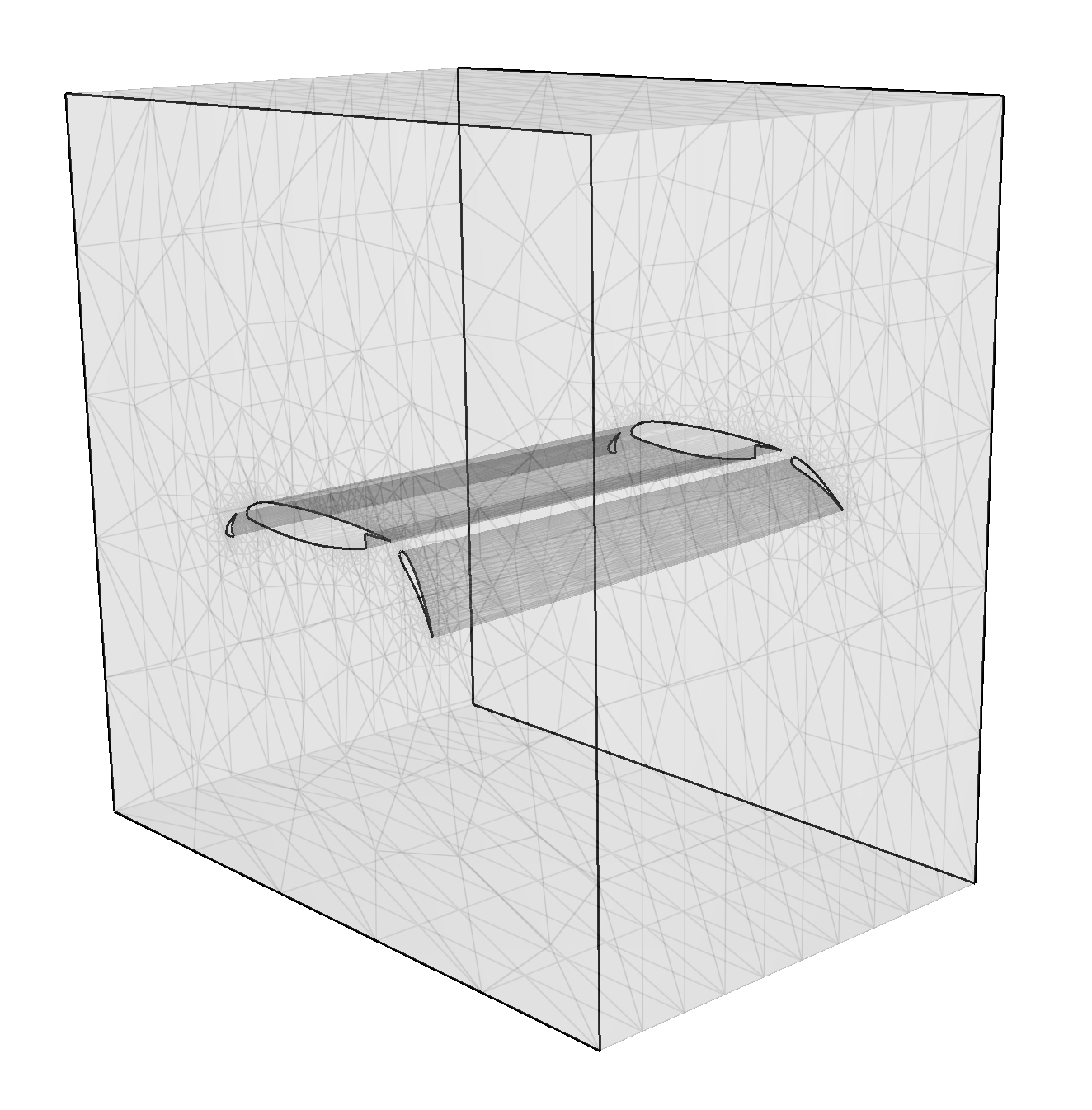}
    \end{subfigure}
    \begin{subfigure}[T]{0.35\textwidth}
      \includegraphics[width=\textwidth]{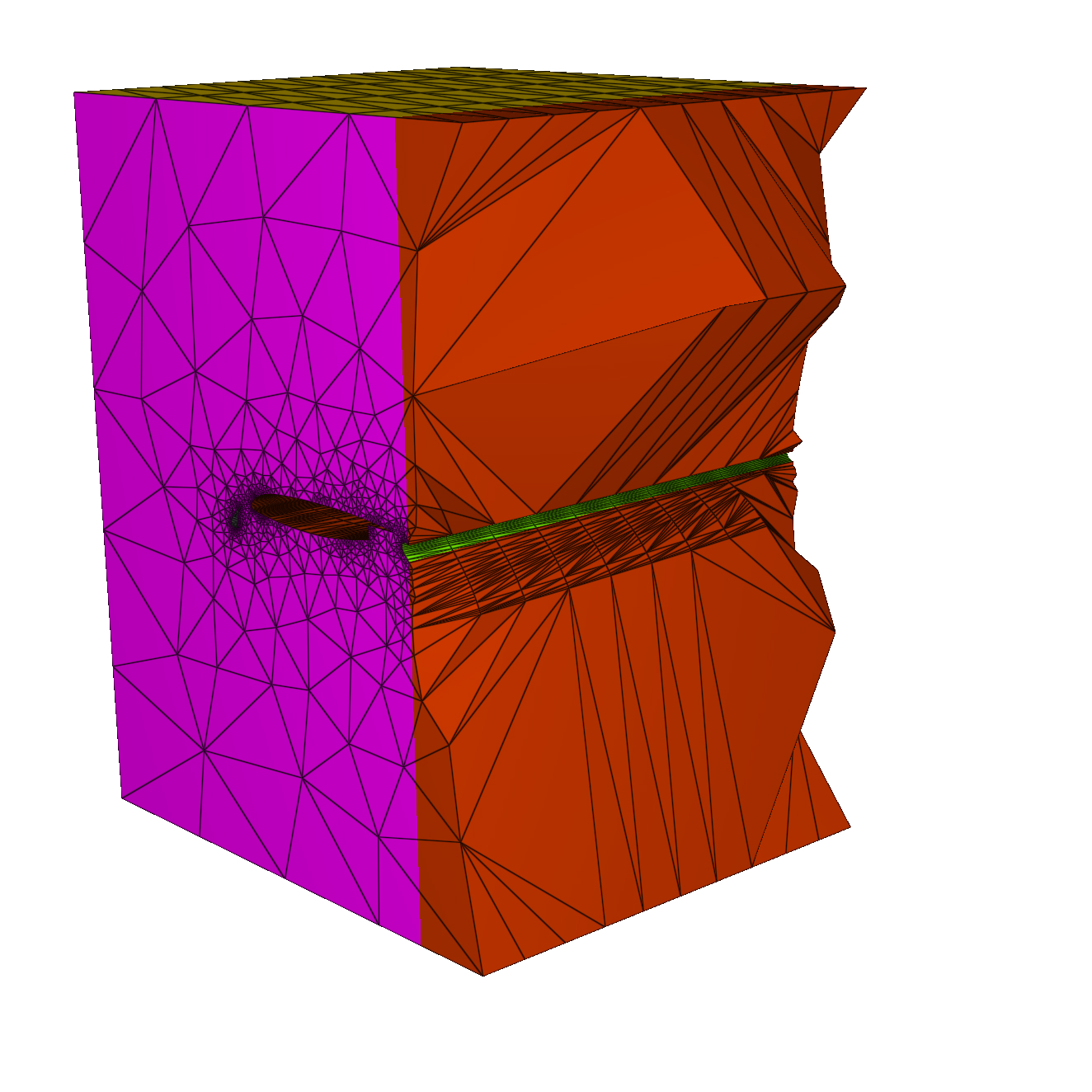}
    \end{subfigure}
    \caption{Illustration of the spacetime meshing problem about a multi-element airfoil. The geometry (left) is a set of curves which trace surfaces when the domain is viewed as a spacetime problem (middle). The flap rotation adds some geometric complexity to the extruded surfaces.
    These surfaces can be triangulated and then passed to a boundary-conforming tetrahedral mesher (right). For three-dimensional spacetime problems, the geometry traces a volume as it moves in the temporal direction, which can be tetrahedralized and then passed to a boundary-conforming pentatope mesher.
    The rightmost mesh was generated with the algorithm described in this paper.}
    \label{fig:spacetime-meshing}
\end{figure*}

Boundary-conforming mesh generation is an essential first step towards running a numerical simulation.
Typically, the bounding surface of the domain is first discretized and then passed as a set of constraints to a volume mesher.
The task of the volume mesher is to produce cells whose faces include the input surface elements.
In two dimensions, this means that input boundary segments must appear as edges of the triangulation; in three dimensions, the input boundary triangles must appear as faces of the tetrahedralization.
In four dimensions, the bounding tetrahedralization must appear as the faces of the pentatopization (a mesh of $4$-simplices, or pentatopes).

Boundary-conforming mesh generation can be categorized as either \emph{constrained} or \emph{conforming}.
Constrained mesh generation consists of exactly preserving the geometry and topology of the input surface.
On the other hand, a conforming mesh may modify the input surface when producing a volume mesh while ideally respecting the input surface as much as possible.
For example, an edge of the input surface may be refined, with the refined sub-edges appearing in the volume mesh.

In two dimensions, the input boundary segments can always be recovered, and a widely used implementation is the \texttt{Triangle} package~\cite{Shewchuk_1996_Triangle}.
In three dimensions, there are well-known polyhedra which do not admit a constrained tetrahedralization~\cite{Chazelle_1984, Schonhardt_1928}.
In such cases, Steiner vertices are \emph{necessary} to produce a boundary-conforming mesh.
\texttt{TetGen}~\cite{Si_2011, Si_2015} is commonly used for producing constrained Delaunay tetrahedralizations, beginning with an unconstrained Delaunay tetrahedralization of the input vertices; \texttt{TetGen} then incrementally recovers the constraining edges and triangles of the boundary.
George et al. also begin with an unconstrained Delaunay tetrahedralization of the input boundary vertices and implement the ``ultimate robustness'' in recovering the boundary edges and triangles~\cite{George_1991, George_2003}.
This robustness is achieved by inserting Steiner vertices at the intersections between the mesh entities and the boundary constraints.
By construction, these refined constraints must ultimately appear as faces of the tetrahedralization.
The authors then eliminate the Steiner vertices by either modifying the topology of the mesh or displacing them into the volume.
Weatherill and Hassan~\cite{Weatherill_1994} use a similar concept but first add field points in the domain interior.
\texttt{CDT3D} also implements the ``ultimate robustness'' approach~\cite{Drakopoulos_2017}; after performing a series of flips to recover as many boundary entities as possible, Steiner vertices are added at the intersections of the boundary constraints with the current mesh entities.

The \texttt{TetWild} algorithm~\cite{Hu_2018, Hu_2020} computes intersection points with a binary space partition and uses exact rational numbers to store vertex coordinates.
This algorithm is very successful, though the runtime is dependent on the envelope size used to control how much detail is preserved in the input.

To date, a complete algorithm for producing boundary-conforming pentatopizations has yet to be demonstrated.
In $4d$, Anderson~\cite{Anderson_2025_PhD} describes the possible intersections between mesh and constraint entities.
Combined with an extension of Weatherill and Hassan's decomposition rules to $4d$, Anderson recovers the boundary of an L-shaped domain in which only segment recovery is needed.
Intersection-based approaches hinge on the robustness of the intersection calculations and can produce many intersection points~\cite{Chen_2017}, making it expensive if used as the sole method for boundary recovery.
Therefore, the goal of this paper is to develop an approach for performing topological operations to recover boundary constraints \emph{before} adding Steiner vertices.

\section{Background}

\subsection{Spacetime Meshing}
Four-dimensional mesh generation is important for supporting spacetime numerical simulations for time-dependent $3d$ problems; see \cref{fig:spacetime-meshing} for an illustration of the boundary-conforming spacetime meshing problem.
Mesh adaptation for such problems has been demonstrated in a tesseract~\cite{Caplan_2019, Caplan_2020, Caplan_2022}, which is the domain traced by a static $3d$ cube.
This consisted of using a local cavity operator to reframe edge splits, collapses, flips and smoothing to adapt the mesh to an anisotropic metric field.
However, more complex domains were not investigated because of an inability to generate initial pentatopizations that respect the domain boundaries.
Recent work in producing boundary tetrahedralizations of moving geometries~\cite{Anderson_2023, Caplan_2025} provide an option for defining the input constraints for the boundary-conforming pentatopization problem.

Behr~\cite{Behr_2008} generated spacetime pentatopizations by discretizing the temporal direction with time slabs and then extrudes an initial spatial mesh at the beginning of one time slab to the end of the time slab, maintaining the connectivity of the spatial mesh while refining the temporal direction.
However, if the domain includes a moving geometry, there is no guarantee that a spatial mesh generated at the beginning of a time slab is a valid spatial mesh representing the domain at the end of the time slab.
For moving geometries, some authors have treated the mesh as an elastic solid, and accounted for topology changes in the moving geometry~\cite{vonDanwitz_2021}.

\subsection{Delaunay Mesh Generation}

Every $d$-simplex in a $d$-dimensional Delaunay mesh contains no other vertex in the interior of its circumscribing sphere.
A classic algorithm for creating (unconstrained) Delaunay meshes is the Bowyer-Watson algorithm~\cite{Watson_1981} which incrementally inserts vertices into a Delaunay mesh by first extracting all simplices violating the empty circumsphere property with respect to the insertion vertex.
This vertex is then connected to the boundary of the cavity which produces valid (positive volume) cells since the Delaunay cavity is \emph{star-shaped}, meaning the insertion vertex is visible to the oriented faces on the boundary of the cavity.

The Delaunay requirement can be extended to the constrained setting by restricting the empty circumsphere property to only those vertices which are \emph{visible} to a boundary constraint.
The constrained Delaunay tetrahedralization of an input set of constraints, represented as a Piecewise Linear Complex (PLC), exists provided the PLC is \emph{ridge}-protected~\cite{Shewchuk_2008}.
Ridges are codimension-2 mesh entities, and a PLC can be determined to be ridge-protected by simply building the unconstrained Delaunay mesh and checking whether every ridge (an edge in $3d$) appears in the mesh.
Shewchuk further discusses gift-wrapping and sweep-based algorithms for producing constrained Delaunay meshes.
Gift-wrapping searches the void on both sides of an individual constraint (a triangle in $3d$) for a vertex to define a constrained cell (a tetrahedron in $3d$).
As noted by Shewchuk, finishing a single face can take $O(n_v n_f)$ time where $n_v$ is the number of vertices and $n_f$ is the number of faces (constraints).
Thus, gift-wrapping is too slow to use in practice.
Shewchuk presents an alternative, which is based on sweeping either a hyperplane or hypersphere~\cite{Shewchuk_2000} but notes this algorithm is difficult to make robust.

The strategy used in \texttt{TetGen}~\cite{Si_2011, Si_2015} is to first construct an unconstrained Delaunay mesh of the input vertices and then recover the constraints using local mesh transformations.
One of \texttt{TetGen}'s strengths is that it minimizes the number of inserted Steiner vertices.
Flips are then used to recover edges, and either flips or a cavity retetrahedralization procedure can be used to insert face constraints~\cite{Si_2013}.
Upon inserting a vertex with a modified Bowyer-Watson procedure, constraint segments and faces deleted by the vertex insertion are then restored using either bistellar flips or a polygon retriangulation procedure. 
However, it is is unclear how such an approach extends to the $4d$ setting.
Furthermore, Diazzi et al.~\cite{Diazzi_2023} have recently pointed out a failure mode in the cavity expansion procedure used in \texttt{TetGen} and propose to implicitly represent Steiner points to improve robustness.

Another option for producing boundary-conforming tetrahedralizations is to augment the input point set such that its unconstrained Delaunay tetrahedralization contains a refined version of the input surface.
The difficulty with such an approach is to ensure the refinement algorithm terminates.
Cohen-Steiner et al.~\cite{CohenSteiner_2004} achieve this using protecting balls of the input constraints, however, the number of vertices increases by a factor of about 3 - 10 for the presented test cases.

\subsection{Advancing Front Approaches}
Advancing front algorithms begin with the boundary surface and either insert cells into an empty domain or into an existing mesh.
This strategy is particularly useful for controlling the mesh spacing near solid boundaries when producing elements suitable for resolving boundary layers~\cite{Lohner_1988, Pirzadeh_1993, Marcum_2001, Peraire_1988, Marcum_2014}.
The former requires costly intersection checks between the proposed cell and the current front, whereas the latter is more robust since front collisions can be detected by marching through the existing mesh.

Advancing-front strategies have also been used for spacetime mesh generation, pioneered by the $2d+t$ tent-pitching algorithm of {\"U}ng{\"o}r and Sheffer~\cite{Ungor_2000} and later extended to higher dimensions~\cite{Erickson_2005}.
The front advances from the boundary at the initial temporal slice by creating cells that satisfy a cone constraint, which is useful for decoupling the system that needs to be solved with discontinuous Galerkin discretizations.

\section{Goals}

The main goal of this paper is to develop an efficient algorithm for recovering as much of the boundary constraints as possible before adding Steiner vertices, with a focus on three- and four-dimensional problems.
While the constrained Delaunay property is well-defined in higher dimensions, the current algorithms for producing constrained Delaunay tetrahedralizations are difficult to extend for producing constrained pentatopizations.
The main idea behind the algorithm is to incrementally advance ridges (codimension-2 entities) into the mesh with a constrained cavity operator.
This approach alleviates the difficulties in scheduling local mesh operations to recover the constraints, particularly in the four-dimensional setting.

Undoubtedly, Steiner vertices are still necessary when this frontal approach stalls.
While the addition of Steiner vertices is not the focus of this paper, the intersection-based method of George et al.~\cite{George_2003} has been implemented to demonstrate that complete boundary conformity is achieved for three-dimensional problems.
An extension of this intersection-based algorithm to four dimensions is also implemented, which enables unstructured boundary-conforming meshes to be generated in $4d$.

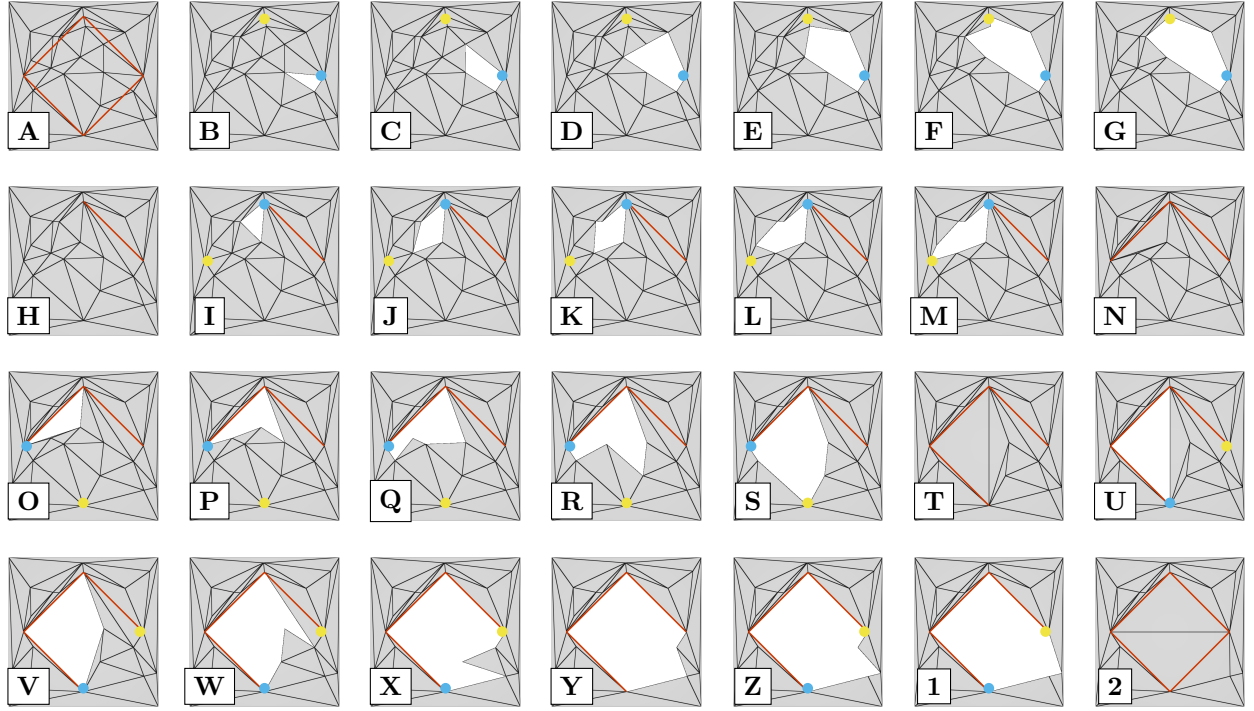
\begin{figure*}[!tb]
  \centering
  \newcommand{\steplabel}[1]{\node [fill=white, draw=black] at (-0.725, -0.725) {\textbf{#1}};}%
  \begin{tikzpicture}
    \node at (0, 0) {\includegraphics[width=0.125\textwidth]{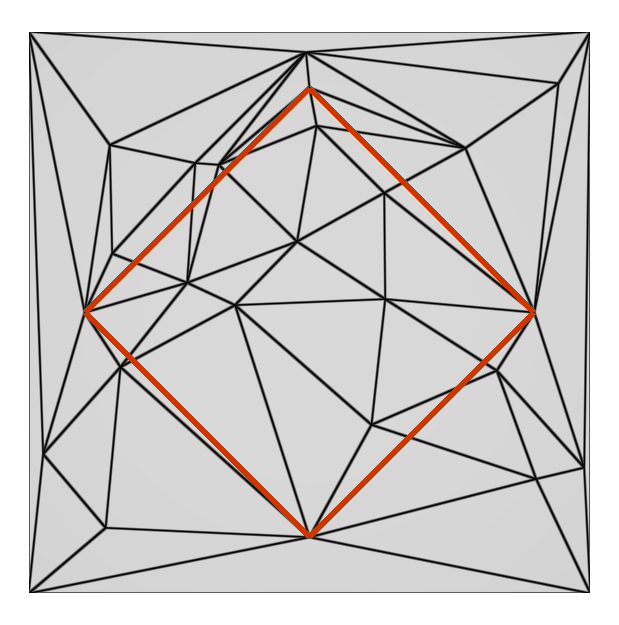}};
    \steplabel{A}
  \end{tikzpicture}%
  \newcommand{\ridge}{\draw[fill=ridgecolor,draw=none] (0.75, 0) circle [radius=2pt];}%
  \newcommand{\constraintnode}{\draw[fill=starcolor,draw=none] (0, 0.75) circle [radius=2pt];}%
  \begin{tikzpicture}
    \node at (0, 0) {\includegraphics[width=0.125\textwidth]{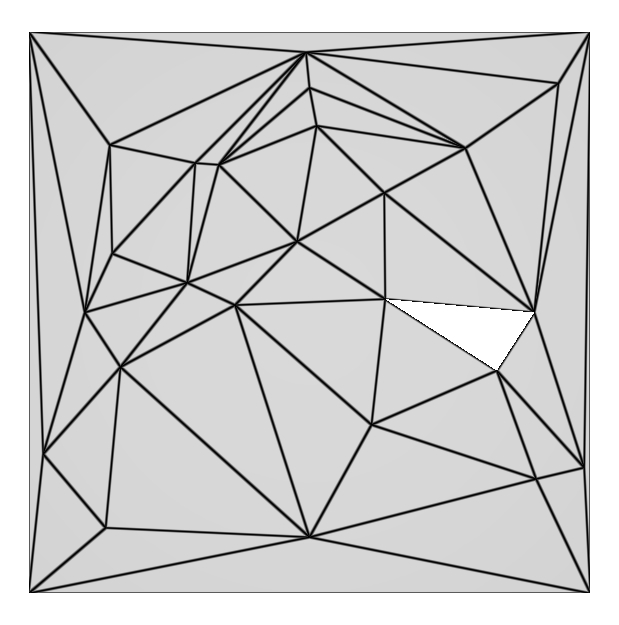}};
    \ridge\constraintnode
    \steplabel{B}
  \end{tikzpicture}%
  \begin{tikzpicture}
    \node at (0, 0) {\includegraphics[width=0.125\textwidth]{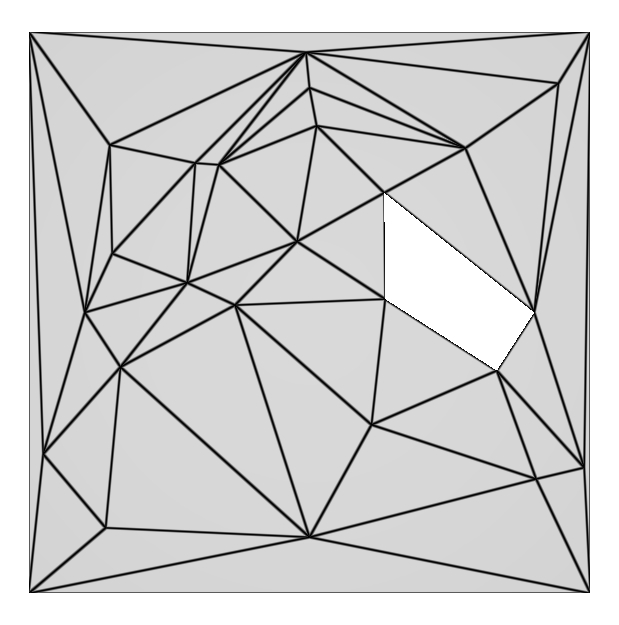}};
    \ridge\constraintnode
    \steplabel{C}
  \end{tikzpicture}%
  \begin{tikzpicture}
    \node at (0, 0) {\includegraphics[width=0.125\textwidth]{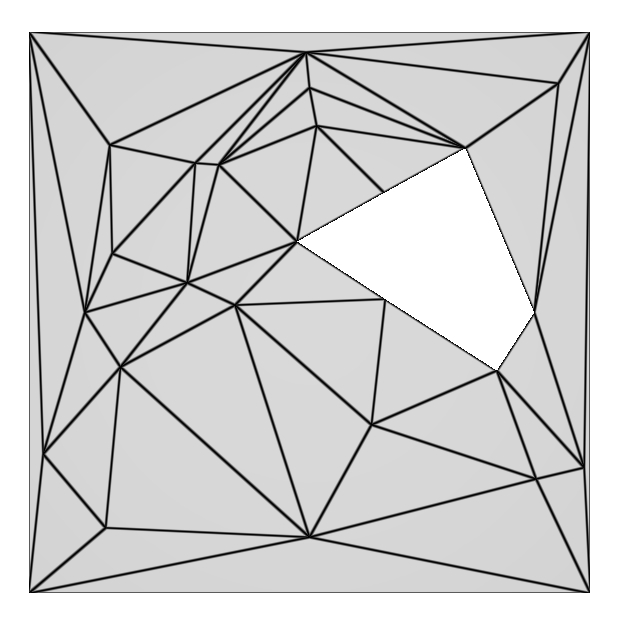}};
    \ridge\constraintnode
    \steplabel{D}
  \end{tikzpicture}%
  \begin{tikzpicture}
    \node at (0, 0) {\includegraphics[width=0.125\textwidth]{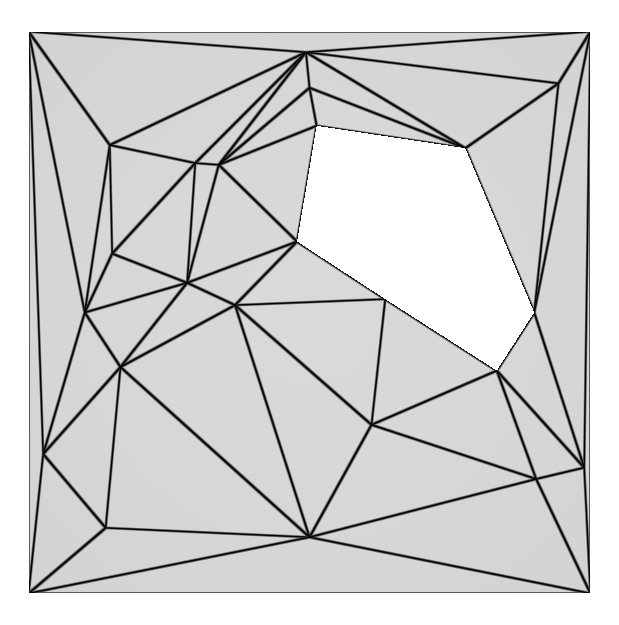}};
    \ridge\constraintnode
    \steplabel{E}
  \end{tikzpicture}%
  \begin{tikzpicture}
    \node at (0, 0) {\includegraphics[width=0.125\textwidth]{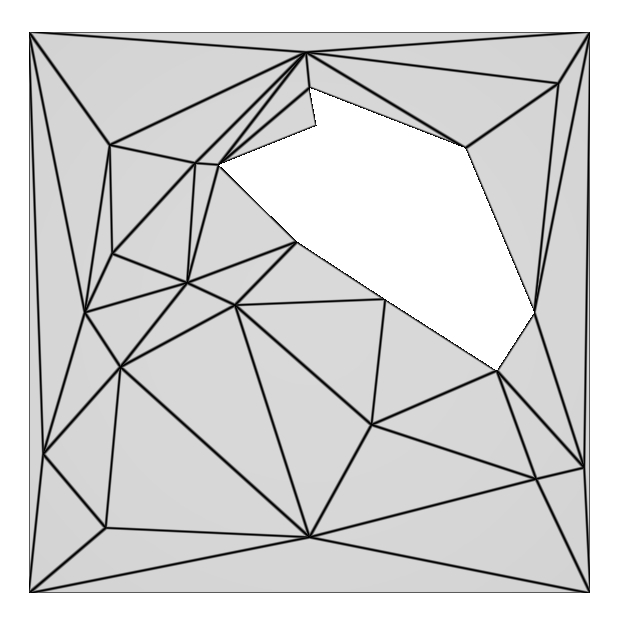}};
    \ridge\constraintnode
    \steplabel{F}
  \end{tikzpicture}%
  \begin{tikzpicture}
    \node at (0, 0) {\includegraphics[width=0.125\textwidth]{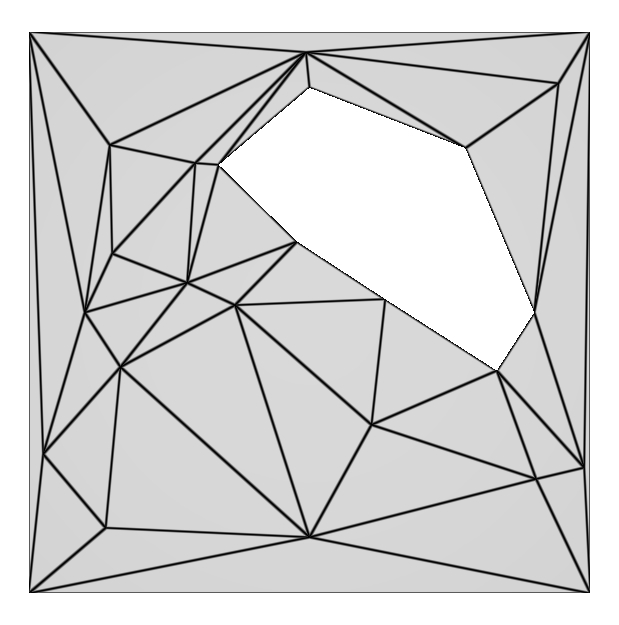}};
    \ridge\constraintnode
    \steplabel{G}
  \end{tikzpicture}%
  \\%
  \begin{tikzpicture}
    \node at (0, 0) {\includegraphics[width=0.125\textwidth]{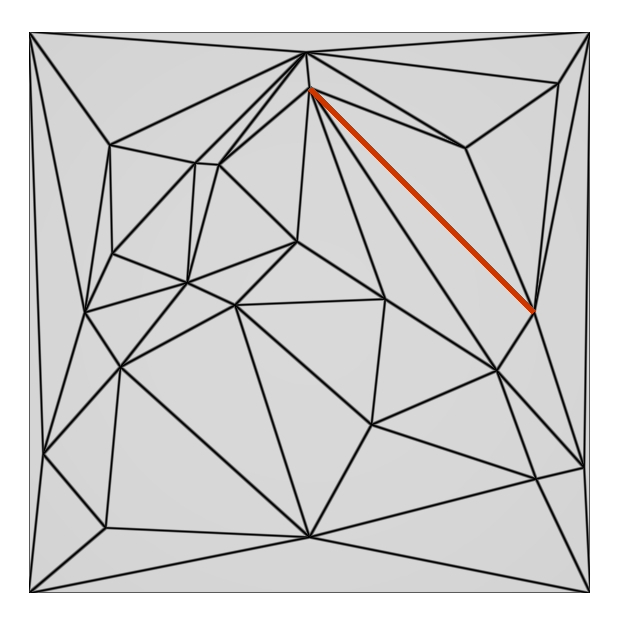}};
    \steplabel{H}
  \end{tikzpicture}%
  \renewcommand{\ridge}{\draw[fill=ridgecolor,draw=none] (0, 0.75) circle [radius=2pt];}%
  \renewcommand{\constraintnode}{\draw[fill=starcolor,draw=none] (-0.75, 0) circle [radius=2pt];}%
  \begin{tikzpicture}
    \node at (0, 0) {\includegraphics[width=0.125\textwidth]{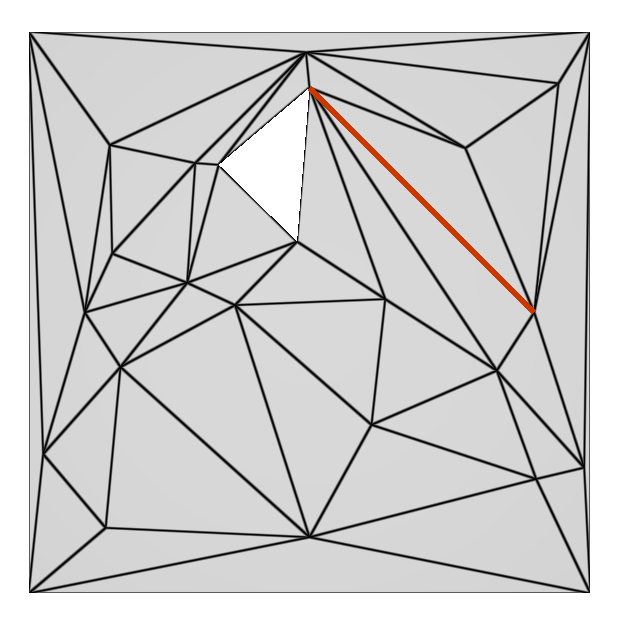}};
    \ridge\constraintnode
    \steplabel{I}
  \end{tikzpicture}%
  \begin{tikzpicture}
    \node at (0, 0) {\includegraphics[width=0.125\textwidth]{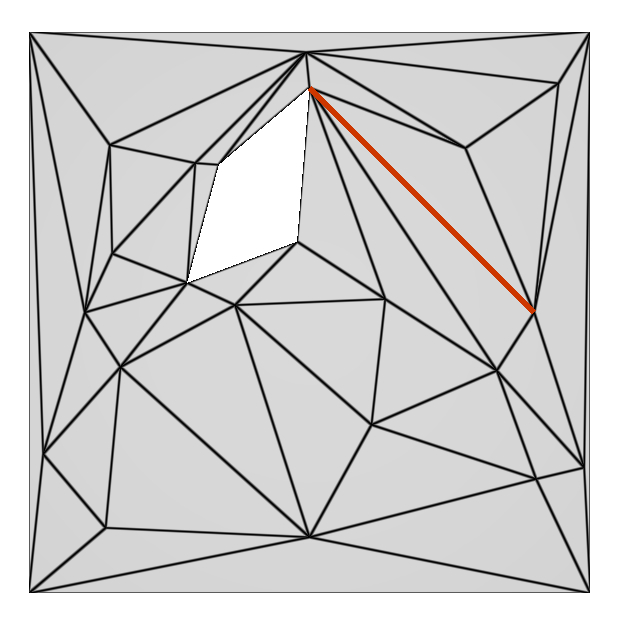}};
    \ridge\constraintnode
    \steplabel{J}
  \end{tikzpicture}%
  \begin{tikzpicture}
    \node at (0, 0) {\includegraphics[width=0.125\textwidth]{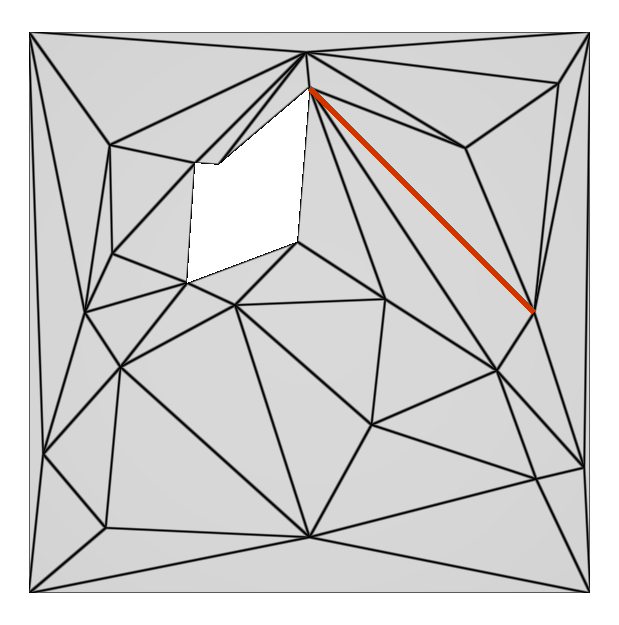}};
    \ridge\constraintnode
    \steplabel{K}
  \end{tikzpicture}%
  \begin{tikzpicture}
    \node at (0, 0) {\includegraphics[width=0.125\textwidth]{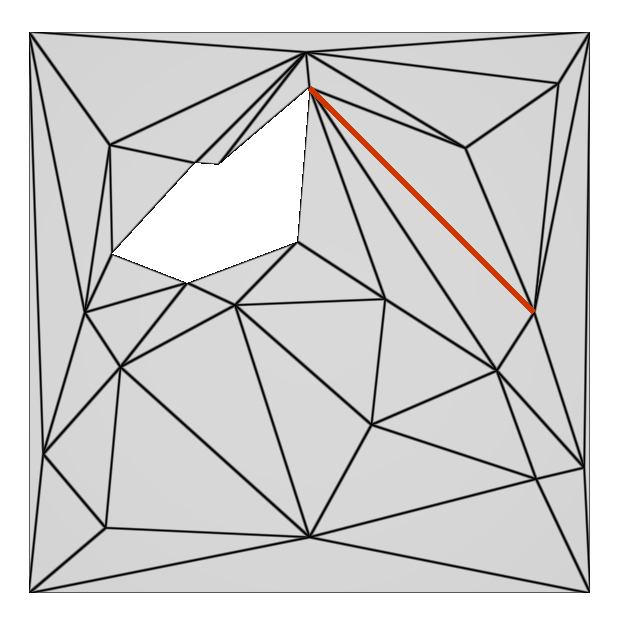}};
    \ridge\constraintnode
    \steplabel{L}
  \end{tikzpicture}%
  \begin{tikzpicture}
    \node at (0, 0) {\includegraphics[width=0.125\textwidth]{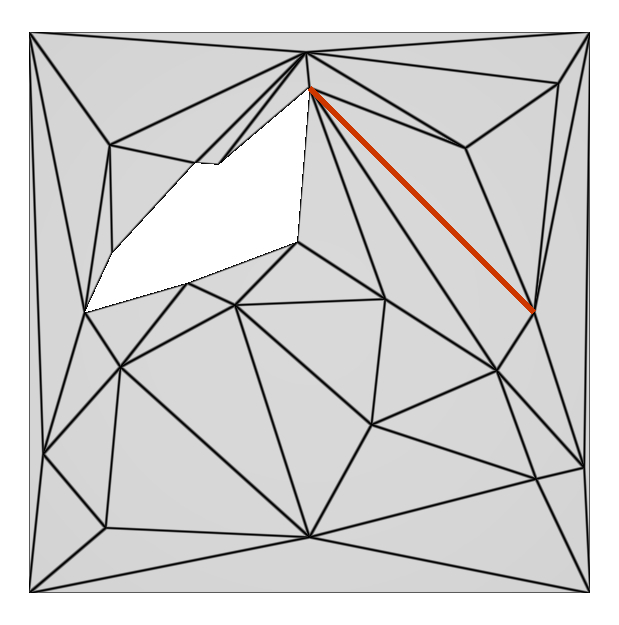}};
    \ridge\constraintnode
    \steplabel{M}
  \end{tikzpicture}%
  \begin{tikzpicture}
    \node at (0, 0) {\includegraphics[width=0.125\textwidth]{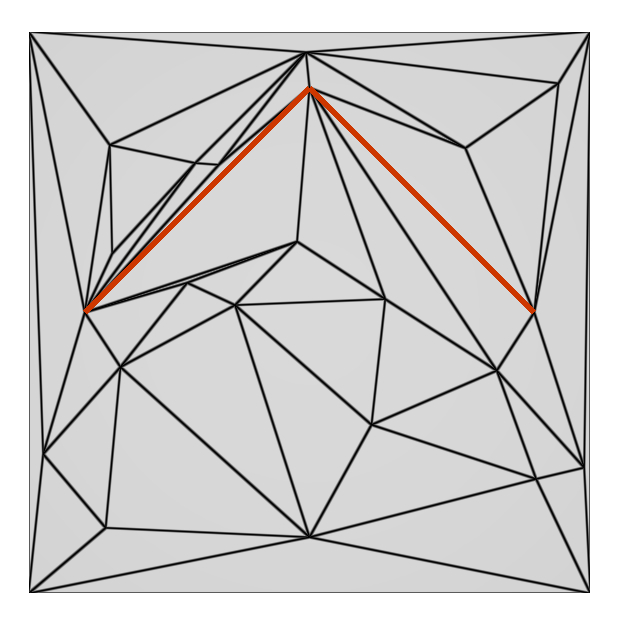}};
    \steplabel{N}
  \end{tikzpicture}%
  \\%
  \renewcommand{\ridge}{\draw[fill=ridgecolor,draw=none] (-0.75, 0) circle [radius=2pt];}%
  \renewcommand{\constraintnode}{\draw[fill=starcolor,draw=none] (0, -0.75) circle [radius=2pt];}%
  \begin{tikzpicture}
    \node at (0, 0) {\includegraphics[width=0.125\textwidth]{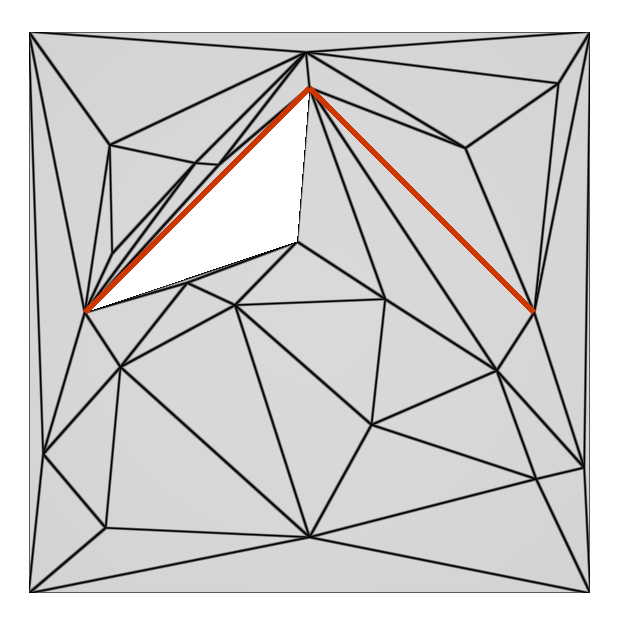}};
    \ridge\constraintnode
    \steplabel{O}
  \end{tikzpicture}%
  \begin{tikzpicture}
    \node at (0, 0) {\includegraphics[width=0.125\textwidth]{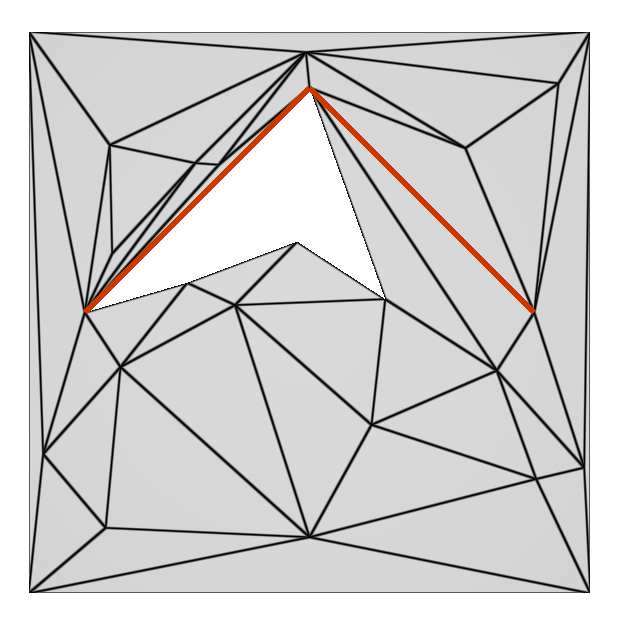}};
    \ridge\constraintnode
    \steplabel{P}
  \end{tikzpicture}%
  \begin{tikzpicture}
    \node at (0, 0) {\includegraphics[width=0.125\textwidth]{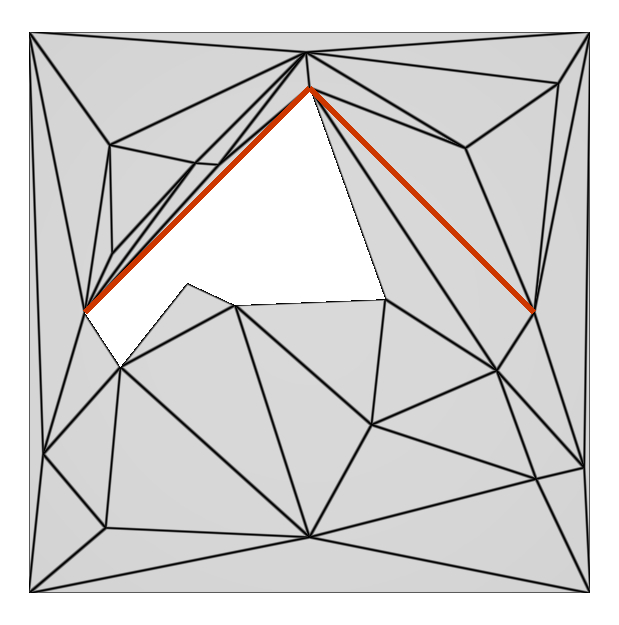}};
    \ridge\constraintnode
    \steplabel{Q}
  \end{tikzpicture}%
  \begin{tikzpicture}
    \node at (0, 0) {\includegraphics[width=0.125\textwidth]{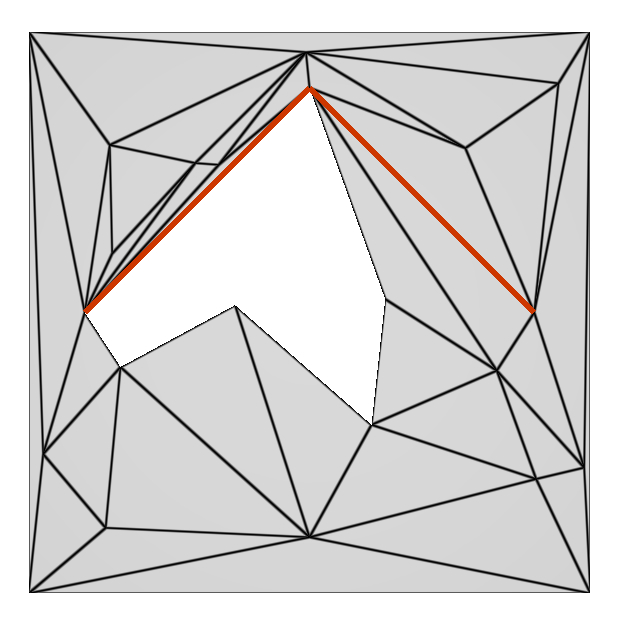}};
    \ridge\constraintnode
    \steplabel{R}
  \end{tikzpicture}%
  \begin{tikzpicture}
    \node at (0, 0) {\includegraphics[width=0.125\textwidth]{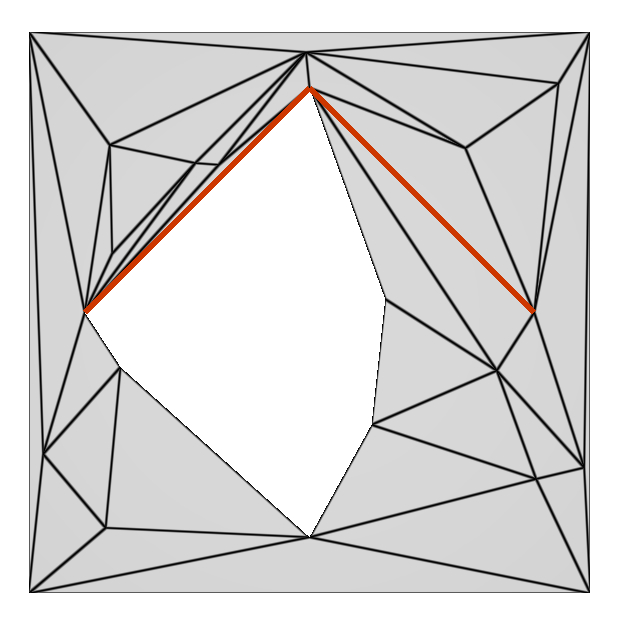}};
    \ridge\constraintnode
    \steplabel{S}
  \end{tikzpicture}%
  \begin{tikzpicture}
    \node at (0, 0) {\includegraphics[width=0.125\textwidth]{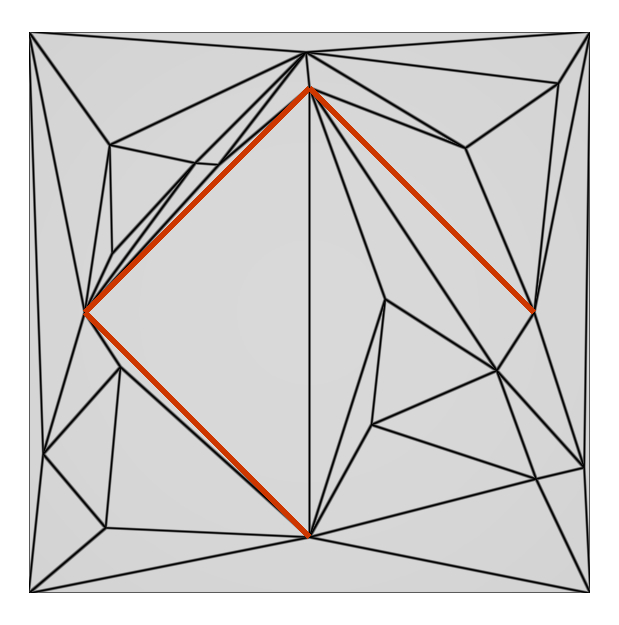}};
    \steplabel{T}
  \end{tikzpicture}%
  \renewcommand{\ridge}{\draw[fill=ridgecolor,draw=none] (0, -0.75) circle [radius=2pt];}%
  \renewcommand{\constraintnode}{\draw[fill=starcolor,draw=none] (0.75, 0) circle [radius=2pt];}%
  \begin{tikzpicture}
    \node at (0, 0) {\includegraphics[width=0.125\textwidth]{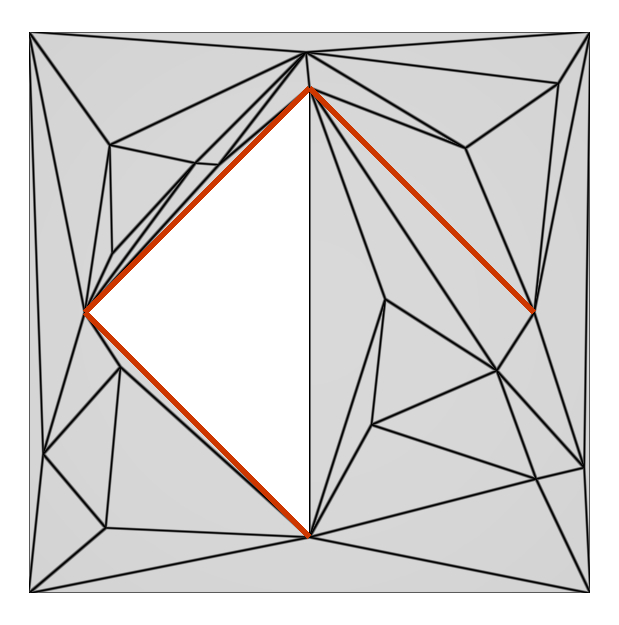}};
    \ridge\constraintnode
    \steplabel{U}
  \end{tikzpicture}%
  \\%
  \begin{tikzpicture}
    \node at (0, 0) {\includegraphics[width=0.125\textwidth]{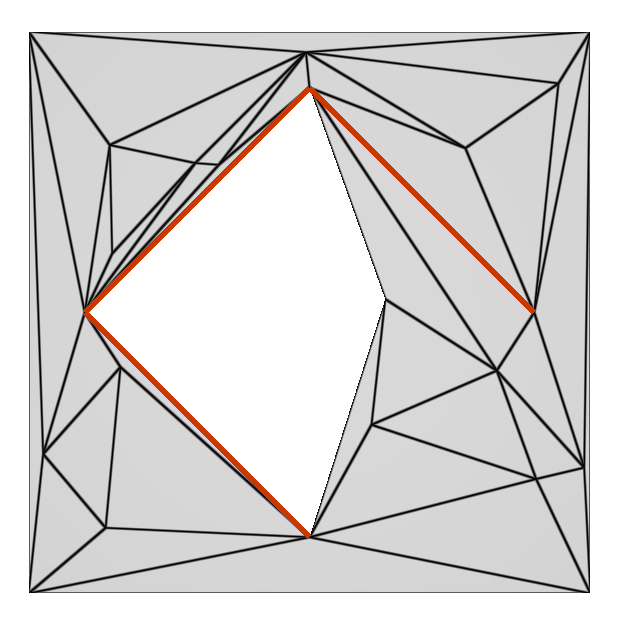}};
    \ridge\constraintnode
    \steplabel{V}
  \end{tikzpicture}%
  \begin{tikzpicture}
    \node at (0, 0) {\includegraphics[width=0.125\textwidth]{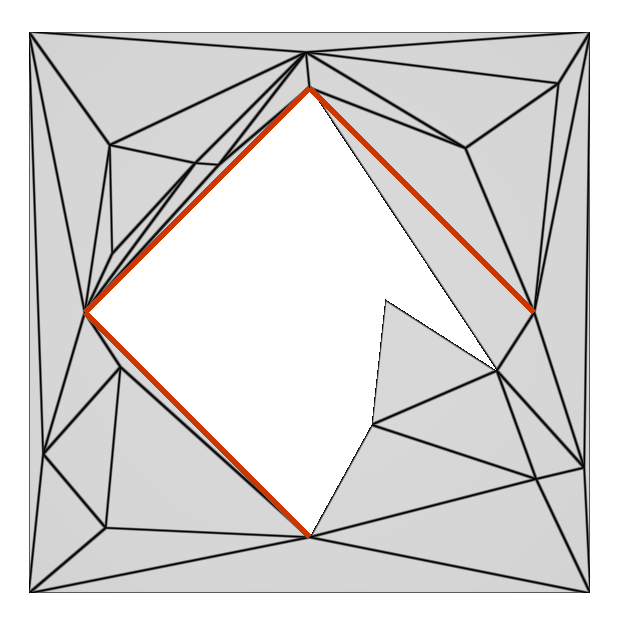}};
    \ridge\constraintnode
    \steplabel{W}
  \end{tikzpicture}%
  \begin{tikzpicture}
    \node at (0, 0) {\includegraphics[width=0.125\textwidth]{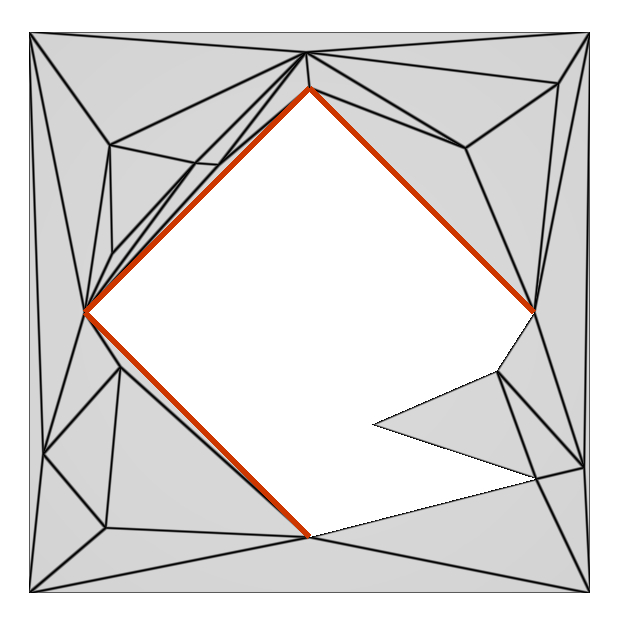}};
    \ridge\constraintnode
    \steplabel{X}
  \end{tikzpicture}%
  \begin{tikzpicture}
    \node at (0, 0) {\includegraphics[width=0.125\textwidth]{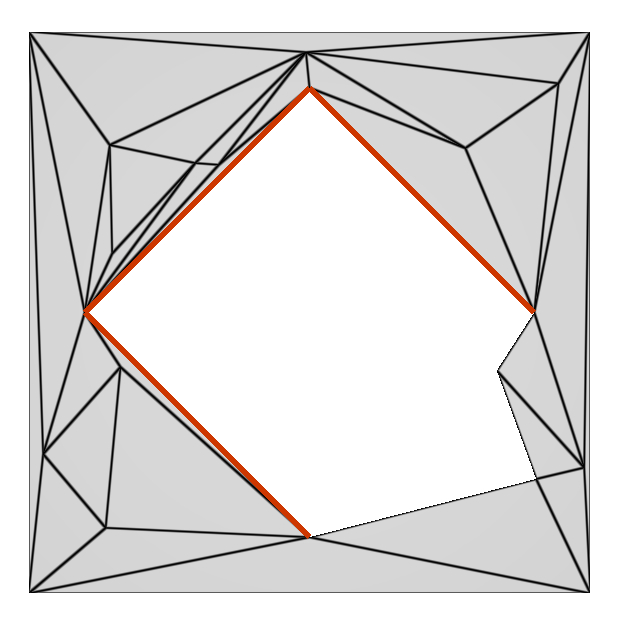}};
    \steplabel{Y}
  \end{tikzpicture}%
  \begin{tikzpicture}
    \node at (0, 0) {\includegraphics[width=0.125\textwidth]{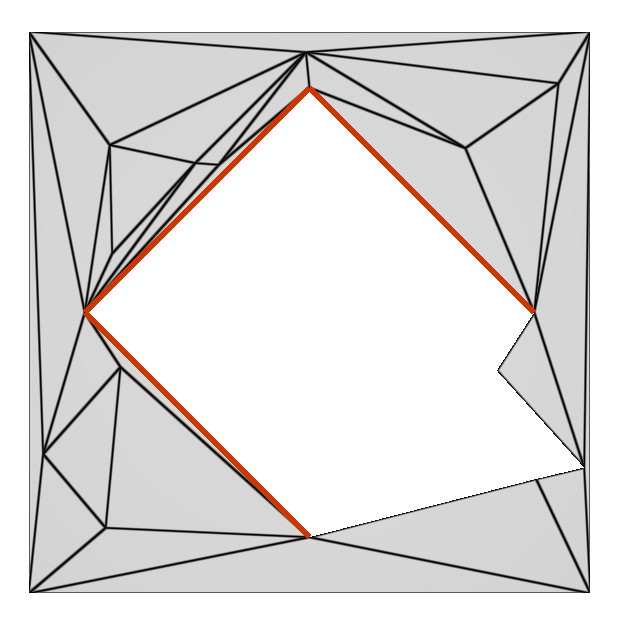}};
    \ridge\constraintnode
    \steplabel{Z}
  \end{tikzpicture}%
  \begin{tikzpicture}
    \node at (0, 0) {\includegraphics[width=0.125\textwidth]{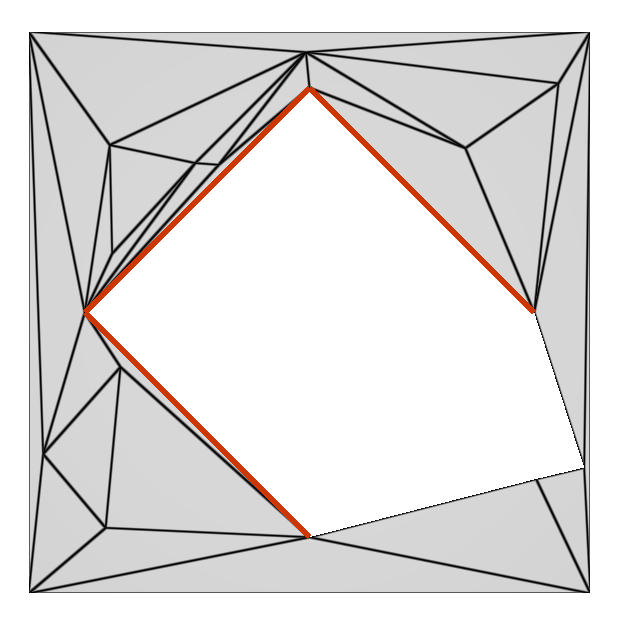}};
    \ridge\constraintnode
    \steplabel{1}
  \end{tikzpicture}%
  \begin{tikzpicture}
    \node at (0, 0) {\includegraphics[width=0.125\textwidth]{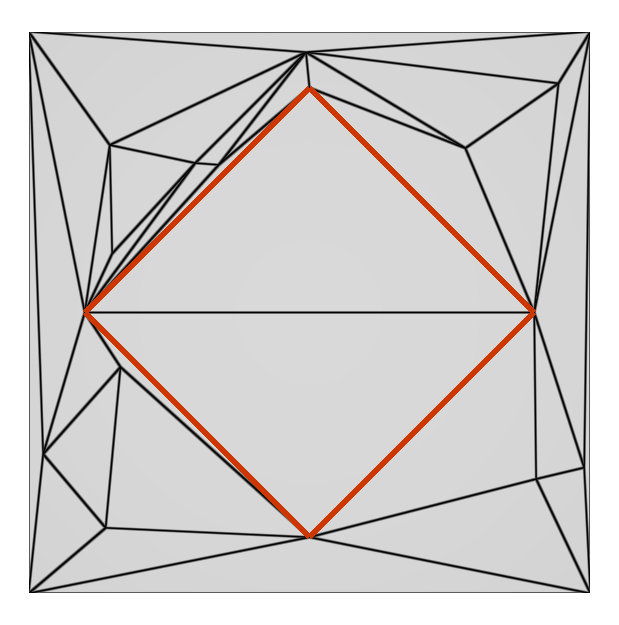}};
    \steplabel{2}
  \end{tikzpicture}%
  \caption{Illustration of the advancing-ridge procedure in $2d$. The initial Delaunay mesh and the four input constraints to recover are shown in subfigure \textbf{A}. The blue dots represent the current front ridge (a vertex in $2d$) and the yellow dots represent the vertex opposite the ridge that defines the constraint to recover. For each constraint to recover, the cavity initializations are shown in subfigures \textbf{B}, \textbf{I}, \textbf{O}, \textbf{U}, and the cavity expansions are illustrated in subfigures \textbf{C}$\to$\textbf{G}, \textbf{J}$\to$\textbf{M}, \textbf{P}$\to$\textbf{S}, \textbf{V}$\to$\textbf{1}. The insertions which recover the constraints are shown in subfigures \textbf{H}, \textbf{N}, \textbf{T}, \textbf{2}.
  }
  \label{fig:insert-constraint}
\end{figure*}
\section{Methodology}
In this work, $\mathcal{B}$ will be used to denote the input $(d-1)$-simplicial mesh of the boundary representing the constraints to be satisfied in the mesh.
The goal is to compute a $d$-simplicial mesh $\mathcal{M}$ such that every constraint appears as a face of some cell in $\mathcal{M}$:
\[
\forall f \in \mathcal{B}, \exists \kappa \in \mathcal{M} \colon f \in \partial\kappa,
\]
where $\partial\kappa$ denotes the set of $d+1$ faces of the $d$-simplex $\kappa$.
The table below summarizes the terminology used in this paper.
The concept of \emph{ridges} is fundamental to the proposed advancing-front technique.

\begin{table}[h]
  \centering
  \begin{tabular}{rllll}
    & Cells & Faces & Ridges \\ \hline
    $2d$ & Triangles & Edges & Nodes \\
    $3d$ & Tetrahedra & Triangles & Edges \\
    $4d$ & Pentatopes & Tetrahedra & Triangles
  \end{tabular}
\end{table}

The notation $\kappa^+(f)$ and $\kappa^-(f)$ refers to the cells on the exterior and interior, respectively, of the oriented mesh face $f$.
Similarly, $f^+(r)$ and $f^-(r)$ refer to the faces on the exterior and interior, respectively, of the oriented ridge $r$ in the boundary mesh.

Combined with an advancing-front technique, the proposed method relies on a constrained version of a cavity operator~\cite{Coupez_2000a, Coupez_2000b, Loseille_2017, Caplan_2019, Caplan_2020, Loseille_2021} to incrementally insert constraints into a mesh.
The cavity operator has been previously used for anisotropic mesh adaptation in $3d$ and $4d$ whereby all classical mesh operations, including edge splits, edge collapses, edge swaps and vertex smoothing are recast as (1) a removal of a set of elements $\mathcal{C}$ from the mesh and (2) a reconnection of the boundary of the cavity ($\partial\mathcal{C}$) to a chosen vertex $p$.
In the generalized cavity operator framework, the topology of the mesh update is described by
\begin{equation}\label{eq:cavity-operator}
  \mathcal{M} = (\mathcal{M} \setminus \mathcal{C}) \cup \lbrace \lbrace p\rbrace \cup f \colon f \in \partial\mathcal{C}\rbrace.
\end{equation}
The two main choices in this framework are (1) the set of elements to remove $\mathcal{C}$ and (2) which reconnection vertex $p$ to select.
This vertex $p$ will sometimes be called the \emph{star} since the cavity should be star-shaped with respect to $p$, meaning $p$ is \emph{visible} to every face of $\partial \mathcal{C}$.
Another way to state this condition is to require that the $d$-simplex formed by connecting $p$ to every $f \in \partial\mathcal{C}$ has a positive volume.

The Delaunay kernel simplifies the two choices to be made in this framework.
In the incremental insertion algorithm of Bowyer and Watson~\cite{Watson_1981}, the vertex being inserted is $p$, and $\mathcal{C}$ is the set of elements whose circumsphere contains $p$.
As long as the initial mesh satisfies the Delaunay property, visibility does not need to be checked since $p$ will always be visible to $\partial\mathcal{C}$.
A $2 \to d$ swap about a face $f$ is also straightforward since $\mathcal{C}$ consists of the two cells adjacent to $f$ and $p$ can be either vertex in the cell opposite the face.
A more generalized $n \to m$ swap results in performing swaps around edges (and ridges in $4d$).
These swaps are essential for producing high-quality meshes but it can be costly to search for the best $p$ in the boundary of $\partial\mathcal{C}$, e.g. when maximizing the minimum metric-based quality over possible insertions~\cite{Caplan_2019}.
Of course, visibility needs to be checked for each insertion candidate.

Loseille first introduced a generalized cavity operator for mesh modification which consisted of expanding the cavity until the re-insertion point $p$ is visible to the cavity boundary~\cite{Loseille_2017}.
He later introduced a constrained version of this operator for inserting quasi-structured tetrahedra that can ultimately be combined to form prisms for boundary layer meshing~\cite{Loseille_2021}.
In this context, tetrahedra formed in a previous layer constrain the cavity operator since they should not be deleted.

\subsection{A frontal approach for inserting constraints with a constrained cavity operator}

Let us now turn to the goal of inserting a single boundary constraint $f \in \mathcal{B}$.
A simple observation about the cavity operator procedure leads to an implementable algorithm.

\begin{observation}
  Given a constraint $f$ to insert into a $d$-dimensional mesh $\mathcal{M}$, if a cavity $\mathcal{C}$ contains some ridge $r$ of $f$ on its exterior, and the missing vertex $p$ where $\lbrace p\rbrace = f\,\backslash\,r$ is either on the cavity interior or exterior, then connecting $p$ to $\partial\mathcal{C}$ will insert $f$ into $\mathcal{M}$.
\end{observation}

\Cref{fig:insert-constraint} and \cref{alg:aro} describe how this observation is used to recover boundary faces.
For clarity, \cref{fig:insert-constraint} is a demonstration of the algorithm in $2d$, despite the focus of the forthcoming results on $3d$ and $4d$.
The same concepts apply, but it should be understood that the depictions of the ridges (in blue) would be edges in $3d$ and triangles in $4d$.
The images should be interpreted from left-to-right, top-to-bottom, with the first image at the top-left showing the initial triangulation of the input points with the four boundary constraints (edges) highlighted in red.
The algorithm can start with any mesh, but a Delaunay mesh of the input vertices is used as the starting point.
Some constraints may already appear as faces of the volume mesh, which are identified and stored in $\mathcal{B}^*$ (\cref{line:extract-boundary}).
It is assumed that $\left|\mathcal{B}^*\right| > 0$, otherwise an initial constraint would need to be inserted to begin the algorithm.
The ultimate goal is to make $\mathcal{B}^* \equiv \mathcal{B}$, though this may not be possible depending on the geometry.
Using the adjacency information of $\mathcal{B}$, the front is then initialized with all ridges incident to a single inserted constraint (\cref{line:initialize-front}).

In each iteration of the algorithm, the blue dots represent the current front ridge being processed, which is popped from the front queue (\cref{line:pop-front}).
The constraint to be inserted only needs to connect the front ridge (in blue) to the yellow vertex.
The cavity is initialized to any cell $\kappa$ that contains $r$, denoted by $R$ (\cref{line:find-root}; \textbf{B}, \textbf{I}, \textbf{O}, \textbf{U}). 
The cavity, beginning with $\kappa$, simply needs to be enlarged until its boundary is visible to the star vertex (depicted in yellow).
In practice, candidate cells $\kappa \in R$ can be further filtered if they are behind the front (remarked upon below).

Starting from some cell $\kappa$ the main task is to construct $\mathcal{C}$ using a series of cavity expansions.
Each iteration of the enlargement procedure scans the boundary faces $\mathcal{F}$ of the current cavity (\cref{line:scan-faces}).
If the insertion vertex $p$ is not visible to a face (\cref{line:not-visible}), then the exterior cell is added to the cavity (\cref{line:expand-cavity-cells}), and the cavity boundary faces are updated (\cref{line:expand-cavity-faces}).
Examples of the expansion procedure for the four constraints in \cref{fig:insert-constraint} are shown in subfigures \textbf{C}~$\to$~\textbf{G}, \textbf{J}~$\to$~\textbf{M}, \textbf{P}~$\to$~\textbf{S} and \textbf{V}~$\to$~\textbf{1}.
The connection with the boundary of the cavity is illustrated in subfigures \textbf{H}, \textbf{N}, \textbf{T} and \textbf{2} (\cref{line:apply-operator}).

In order to ensure each step of the algorithm does not regress from the ultimate goal of recovering the boundary, constraints that were previously inserted ($\mathcal{B}^*$) should remain in the mesh.
This adds an additional verification step to the cavity expansion procedure: when encountering a constraint that was previously inserted, if $p$ is not visible to the corresponding face, then the cavity cannot be enlarged, and the insertion of the constraint must be abandoned (\cref{line:constrained-cavity}).
When this happens, it may be that the topology of the mesh is sufficiently modified by future insertions to allow it to succeed on a later attempt.
As a result, the constraint is requeued for another attempt (\cref{line:re-add-to-front}).
A maximum of $d$ attempts are made for each constraint, which is inspired by the fact that the constraint may be inserted from any of its $d$ ridges that may appear on the front.

For brevity, this Advancing-Ridge cavity Operator algorithm will now be referred to as ARO.

\newpage
\begin{algorithm}[Advancing-Ridge Operator]{}\quad
  \algrenewcommand{\algorithmiccomment}[1]{\ \(\triangleright\) #1}
  \begin{algorithmic}[1]
  \Require{Closed, manifold boundary mesh $\mathcal{B}$}
  \Require{Initial mesh $\mathcal{M}$ of the input vertices}
  \Common{$\mathcal{M}$, $\mathcal{B}^* \subseteq \mathcal{B}$}
    \State $\mathcal{F} \gets \bigcup_{\kappa \in\mathcal{M}}\partial \kappa$ \Comment{All faces in $\mathcal{M}$}
    \State $\mathcal{B}^* \gets \mathcal{B} \cap \mathcal{F}$\Comment{Extract constraints already in $\mathcal{M}$}\label{line:extract-boundary}

    \State \hspace{-2pt}\Comment{Initialize the front}
    \State $Q \gets \lbrace r  \colon f^-(r) \in \mathcal{B}^* \land f^+(r) \notin\mathcal{B}^*\rbrace$\label{line:initialize-front}

    \While{$\left|Q\right| > 0$}
      \State $r \gets \mathrm{Pop}(Q)$ \Comment{Get next front ridge}\label{line:pop-front}
      \State $f \gets f^+(r)$
      \If{$\mathrm{InsertConstraint}(r, f)$}
        \For{$r_f \in \partial f \setminus \lbrace r\rbrace$}
          \State $Q \gets Q \cup \lbrace r_f\rbrace$ \label{line:update-front}
        \EndFor
      \Else
        \State $Q \gets Q \cup \lbrace r\rbrace$ \Comment{If \# attempts $\le d$}\hfill\label{line:re-add-to-front}
      \EndIf
    \EndWhile\label{line:end-front}

    \Function{InsertConstraint}{$r$, $f$}
      \State $\lbrace p\rbrace \gets f \setminus r$
      \State $R \gets \lbrace \kappa \in \mathcal{M} \colon r \subset \kappa\rbrace$\label{line:get-ring}
      \If{$\exists \kappa \in R \colon f \in \partial\kappa$}
        \State \Return \texttt{True} \Comment{$f$ is already in $\mathcal{M}$}\hfill\label{line:constraint-already-exists}
      \EndIf
      \For{$\kappa \in R$} \label{line:find-root}
          \State $(\mathcal{C}, \partial\mathcal{C}) \gets \mathrm{BuildVisibleCavity}(\kappa, p)$
          \If{$\mathcal{C} \neq \emptyset$}
            \State Apply \cref{eq:cavity-operator}.\label{line:apply-operator}
            \State \Return \texttt{True}
          \EndIf
      \EndFor
      \State \Return \texttt{False}
    \EndFunction

    \Function{BuildVisibleCavity}{$\kappa$, $p$}
      \State $\mathcal{C} \gets \lbrace\kappa\rbrace$
      \State $\partial\mathcal{C} \gets \partial\kappa$ \Comment{Cavity boundary}
      \Repeat
        \State $\texttt{visible} \gets \texttt{True}$
        \For{$f \in \mathcal{\partial C}$}\label{line:scan-faces}
          \If{$\mathrm{Volume}(f \cup \lbrace p\rbrace) < 0$} \label{line:not-visible}
            \If{$f \in \mathcal{B}^*$}
              \State\hspace{-2pt}\Comment{Do not undo constraints}
              \State\Return$(\emptyset, \emptyset)$\label{line:constrained-cavity}
            \EndIf
            \State $\texttt{visible} \gets \texttt{False}$
            \State $\mathcal{C} \gets \mathcal{C} \cup \kappa^+(f)$ \Comment{Add exterior cell} \label{line:expand-cavity-cells}
            \State $\mathcal{\partial C} \gets (\partial\mathcal{C}  \cup \partial\kappa^+(f)) \setminus \lbrace f\rbrace$\label{line:expand-cavity-faces}
          \EndIf
        \EndFor
      \Until{\texttt{visible}}
      \State \Return $(\mathcal{C}, \partial\mathcal{C})$
    \EndFunction
  \end{algorithmic}
  \label{alg:aro}
\end{algorithm}%
\subsection*{Remarks}
\begin{enumerate}
  \item Clearly, if $\mathcal{B}^*$ is initially empty (\cref{line:extract-boundary}), then the algorithm cannot advance. For all the cases studied in this paper, there is always at least one face of the initial (unconstrained) Delaunay mesh which is an element of $\mathcal{B}$. However, it is possible this is not the case, and the algorithm would need to insert some constraint to initialize the front.
  \item Candidate starting cells $\kappa$ of the cavity expansion (\cref{line:find-root}) can be filtered using the geometry of the cells. Regardless of the dimension, there are always \textbf{two} vertices $u$, $v$ such that $u \in \kappa$, $v \in \kappa$ opposite the ridge ($u \notin r$, $v \notin r$). The cells can be filtered by requiring that $u$ and $v$ are on opposite sides of the plane defined by the constraint to be inserted.
  \item Exact geometric predicates are used to compute the volume in \cref{line:not-visible}. An \texttt{orient4d} predicate was developed using the Predicate Construction Kit~\cite{Levy_2016}.
  \item The overall approach performs several iterations of the frontal insertion procedure of \cref{alg:aro}. While most constraint insertions occur on the first pass, $2 \to d$ swaps are also used to improve the cell mean ratios, which has the potential to ``unlock'' subsequent insertions.
\end{enumerate}

\subsection{Design of the cavity operator implementation}

Certain components of \cref{alg:aro} can be made very efficient by carefully designing the cavity operator.
In particular, the way cell-to-cell (in $\mathcal{M}$) and face-to-face (in $\mathcal{B}$) adjacencies are retrieved and updated is critical for efficiently expanding the cavity and applying the insertion.
The design of the adjacency storage mostly follows Marot's proposal for tetrahedral meshes~\cite{Marot_2018}, whereby (1) a single integer encodes the cell and face index and (2) the adjacency information is mirrored between adjacent cells.
Specifically, this scheme enforces the adjacency array \texttt{adj} to satisfy:

\begin{equation}
\texttt{adj}(a_i) = a_j \iff \texttt{adj}(a_j) = a_i,
\end{equation}

where $a_i$ and $a_j$ are the encoded cell-face integers of two cells sharing a face.
The main idea is to reserve a few bits at the beginning of integers $a_i$ and $a_j$ which contain the index of the corresponding face.
For simplicial meshes, this face index $i$ refers to the index of the face opposite vertex $i$ in the simplex.
A simplex of dimension $d$ has $d + 1$ faces, so these bits must be capable of representing $d+1$ numbers.
For $2d$ and $3d$ meshes, two bits can represent the required face indices but three bits are needed in $4d$.
With a 64-bit representation of the adjacencies, this has little impact on the maximum number of cells that can be represented since 61 bits are still available to represent the cell indices.

Bitwise operators can be used to efficiently encode and decode the adjacency information.
The encoding of the adjacency information $a$ across face $i$ of element $k$ is $a = k\,\verb!<<!\,b_d + i$, where $b_d$ is the dimension-dependent number of bits reserved for the face index: $b_2 = 2$, $b_3 = 2$ and $b_4 = 3$.
Given an adjacency $a$, the corresponding element $k$ is therefore $k = a\,\verb!>>!\,b_d$ and the corresponding face index is $i = a\,\verb!&!\, 2^{b_d - 1}$.
This scheme enables an efficient access and update of the functions used in \cref{alg:aro}, notably $\kappa^+(f)$, $\kappa^-(f)$, $f^+(r)$ and $f^-(r)$.
The use of \texttt{C++} templates significantly simplifies the dimension-independent implementation of these functions.

Marot also proposes an efficient memory-aligned point data structure that uses 3 64-bit \texttt{double}s to represent the coordinates, and an additional 64-bit padding variable to store temporary vertex data~\cite{Marot_2018}.
This ensures each vertex consumes 32 bytes and will be less likely to overlap multiple cache lines during a memory fetch.
A padding variable is also used here to store additional data for each vertex, but it becomes difficult to enforce the same memory alignment pattern in $4d$ since the 4 \texttt{double} coordinates for each vertex already consume 32 bytes.
Instead of trying to align the vertex structure size to a typical cache line size, the padding is kept to a 64-bit variable which stores auxiliary data such as thread index and the index of a single simplex referencing the vertex.
The latter can be used as a starting point for traversing the adjacencies when retrieving all cells referencing a vertex, which can also be used to compute $R$ in \cref{line:get-ring} of \cref{alg:aro}.

Similar to Marot's suggestion, a lookup table is used to compute the internal adjacencies of the created cells of \cref{eq:cavity-operator}.
The size of the lookup table in $4d$ was chosen to be $256\times256\times256$ which means it can only accommodate $256$ vertices in the cavity boundary.
When there are too many vertices on the cavity boundary, the implementation resorts to a map for associating the cavity boundary ridges with the adjacency information.
In $3d$, it is possible to estimate when this happens using the Euler characteristic, but no such relationship was found for $4d$.
Instead, the number of vertices is estimated to be $\left|\mathcal{F}\right|/4$ and if this is breached, the implementation resorts to the map-based approach to complete the insertion.
\subsubsection{Performance assessment.}
\label{sec:performance}
Before presenting some results related to \cref{alg:aro}, the performance of the cavity operator implementation will be evaluated by generating a series of unconstrained Delaunay meshes in $2d$, $3d$ and $4d$.
While the operator has also been designed to be thread-safe, the following analysis is only done for the sequential implementation since the parallel scheduling and partitioning of the cavity operator is preliminary and only achieves a parallel speedup factor of 2 (on average) with 10 threads.
The following results were obtained using a workstation laptop with 22 Intel Ultra 9 185H cores and 64 GB of RAM and the \texttt{C++} implementation was compiled with \texttt{GCC} 13.3.0.

\begin{figure}[h!]
  \centering
  \includegraphics[width=0.5\textwidth]{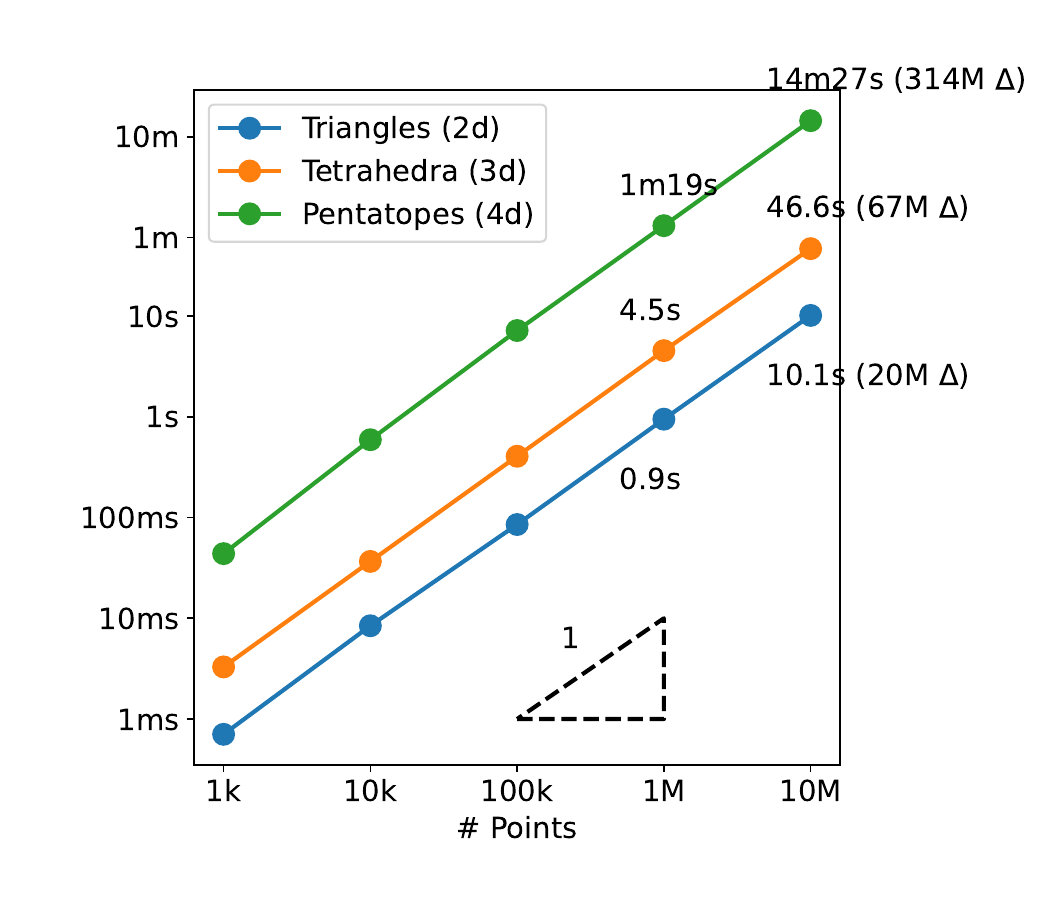}\\
  \includegraphics[width=0.5\textwidth]{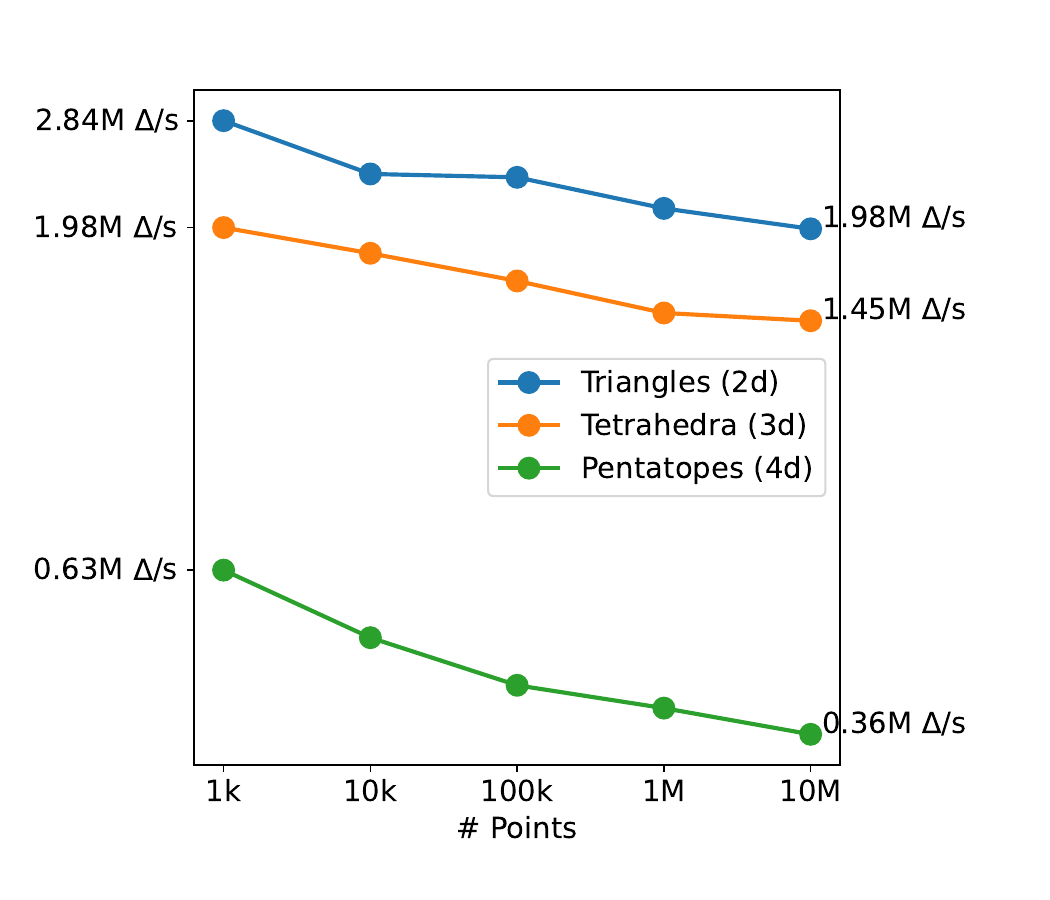}
  \caption{With a single thread, the current cavity operator implementation can insert 20 million triangles ($2d$) in 10 seconds, 67 million tetrahedra ($3d$) in 46.6 seconds and 314 million pentatopes ($4d$) in 14 minutes and 27 seconds. $\Delta/\mathrm{s}$ refers to the simplex creation rate in simplices per second.}
  \label{fig:performance}
\end{figure}

\Cref{fig:performance} shows the total time to insert up to 10 million randomly distributed vertices with the Delaunay kernel in $2d$, $3d$ and $4d$.
The vertices are first sorted along a Hilbert curve before inserting them into the mesh.
In $2d$, 20 million triangles were created in 10.1 seconds, 67 million tetrahedra were created in 46.6 seconds ($3d$) and 314 million pentatopes were created in 14 minutes and 27 seconds ($4d$).
The simplex creation rate (measured as number of simplices divided by total time) decreases slightly as the mesh size grows, and is generally slower as the mesh dimension increases.
Nonetheless, for the largest mesh, a creation rate of about 360,000 pentatopes per second is very reasonable.

To further situate the speed of this dimension-independent cavity operator implementation, a $3d$ Delaunay tetrahedralization with 5 million vertices was generated in 22.5 seconds.
At the beginning of Marot's paper, it is reported that their implementation inserts 5 million vertices in 12.7 seconds.
While the current implementation is slower than this, it is nonetheless faster than what Marot reports for other implementations (\texttt{geogram}: 34.6 seconds; \texttt{TetGen}: 32.9 seconds; \texttt{CGAL}: 33.8 seconds).

\section{Experiments}
Various test cases will now demonstrate the ability of the ARO algorithm to recover much of the input boundary mesh.
As described earlier, the cavity might not be able to expand without stepping across a previously recovered constraint and should, therefore, be abandoned.
After the advancing-ridge procedure stalls, intersections between mesh entities and constraints ahead of the stalled front ridges are calculated and inserted into $\mathcal{B}$, thus subdividing some of these constraints.
The updated $\mathcal{B}$ is then used for the next iteration of the ARO algorithm.
Generally, this procedure terminates in a few iterations.

While this Steiner vertex calculation is not the focus of the paper, it is worth mentioning that for $3d$ boundary recovery, edge-edge and edge-face intersections are calculated and, in $4d$, edge-edge, edge-triangle, edge-tetrahedron and triangle-triangle intersections are calculated.
Care should be taken for intersections near entity boundaries (e.g. when an intersection with a mesh triangle occurs near one of its edges).

Similar to \Cref{sec:performance}, timing was measured on a workstation laptop with 22 Intel Ultra 9 185H cores and 64 GB of RAM.

\subsection{Boundary-conforming~tetrahedralizations.}
The following examples demonstrate the ability of the ARO algorithm to produce boundary-conforming tetrahedralizations.
The test cases include meshes of the Stanford Bunny~\cite{Turk_1994}, Keenan Crane's 3D Model Repository~\cite{Crane_2013} as well as models of a tire, bottle, lander~\cite{Haimes_2013_ESP}, and the Falcon aircraft~\cite{Loseille_2018}.
The method also succeeded on boundary meshes with stretched elements, such as the mesh of a giraffe~\footnote{\tiny"Low Poly Giraffe" (\url{https://skfb.ly/6WBsx}) by W3DGEx is licensed under Creative Commons Attribution (http://creativecommons.org/licenses/by/4.0/).} and the initial anisotropic curvature-adapted mesh of the ONERA-M6 wing~\cite{Balan_2020}.
It also works when the boundary has multiple connected components, which was tested with a surface triangulation generated for the simplified high-lift Common Research Model (CRM-HLS) of the sixth High-Lift Predication Workshop. 
The surface mesh for the latter geometry, in addition to the tire, bottle and lander~\cite{Haimes_2013_ESP} were generated using an in-house isotropic surface remesher that adapts to the local curvature of the geometry.

\begin{figure*}[h!]%
  \centering%
  \begin{subfigure}{0.175\textwidth}%
    \includegraphics[height=2.5cm]{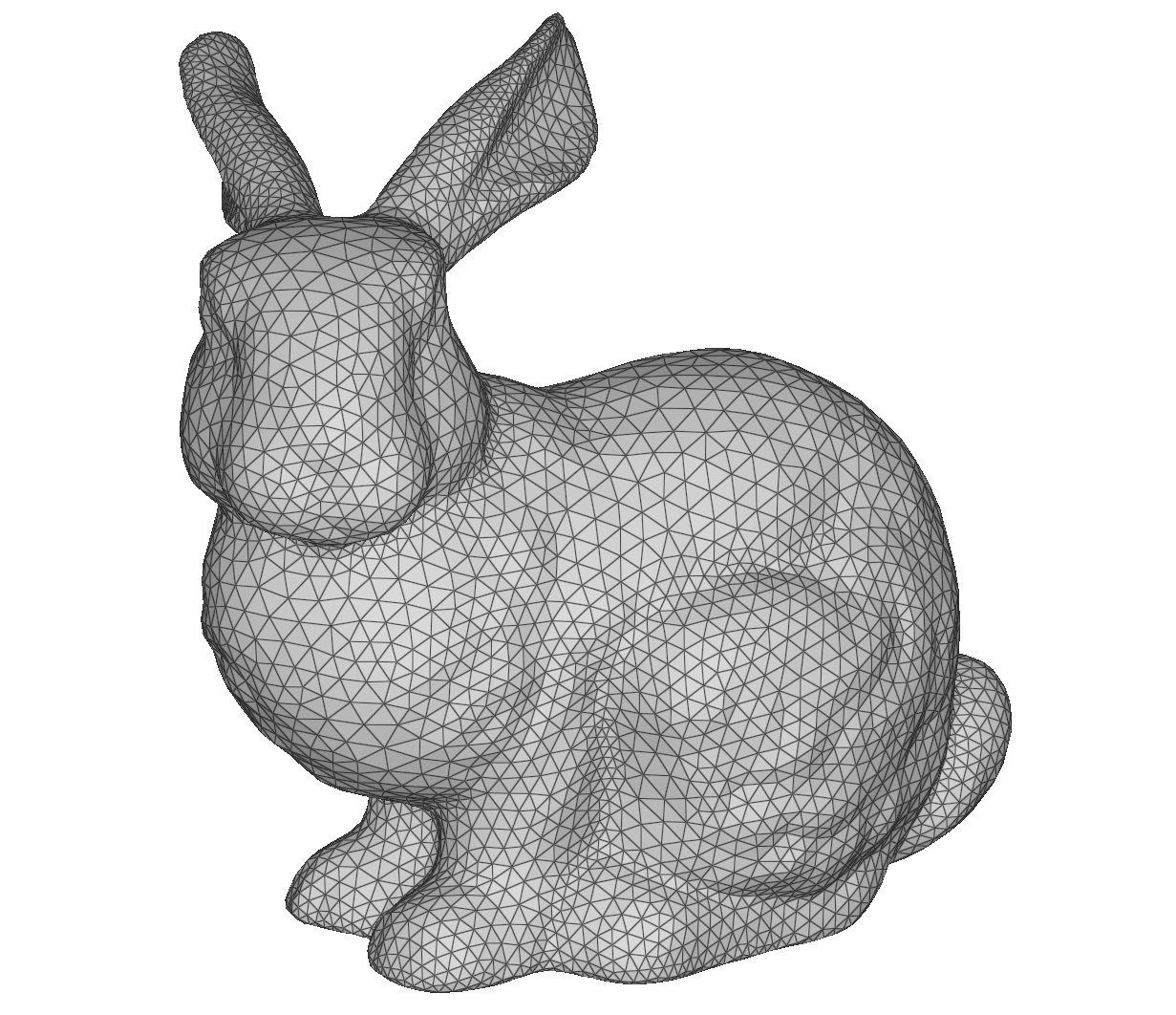}%
    \caption{Stanford Bunny}
  \end{subfigure}\hfill%
  \begin{subfigure}{0.15\textwidth}%
    \includegraphics[height=2.5cm]{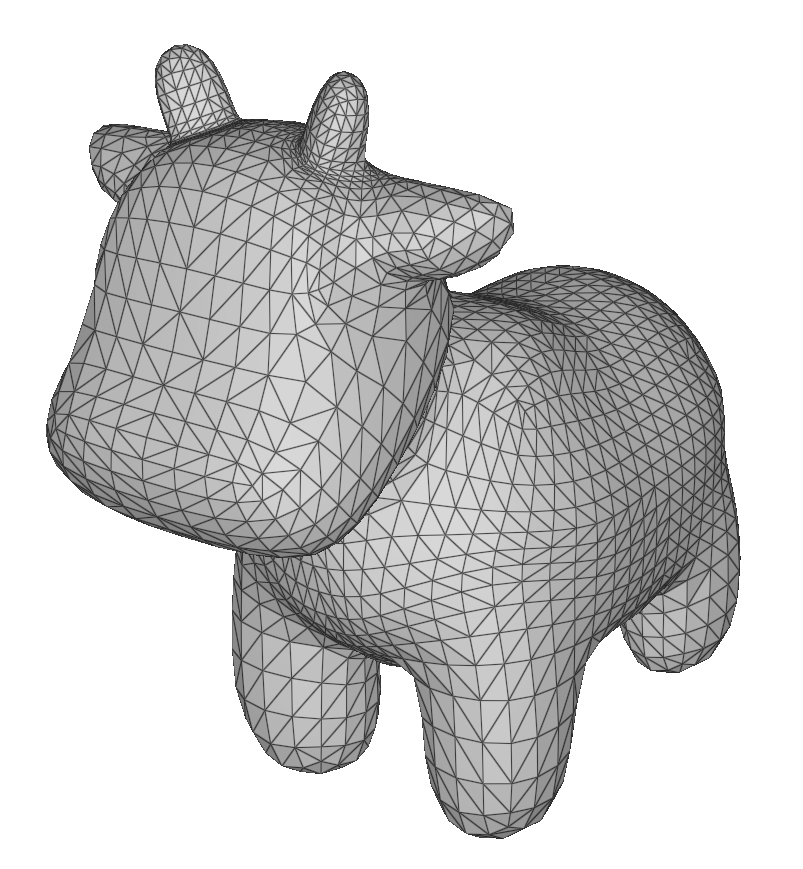}%
    \caption{Spot}
  \end{subfigure}\hfill%
  \begin{subfigure}{0.265\textwidth}%
    \includegraphics[height=2.5cm]{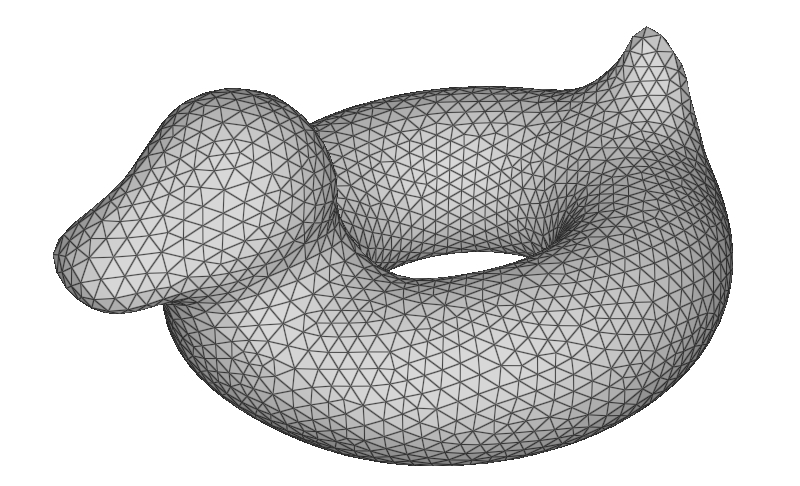}%
    \caption{Pool Toy}
  \end{subfigure}\hfill%
  \begin{subfigure}{0.15\textwidth}%
    \includegraphics[height=2.5cm]{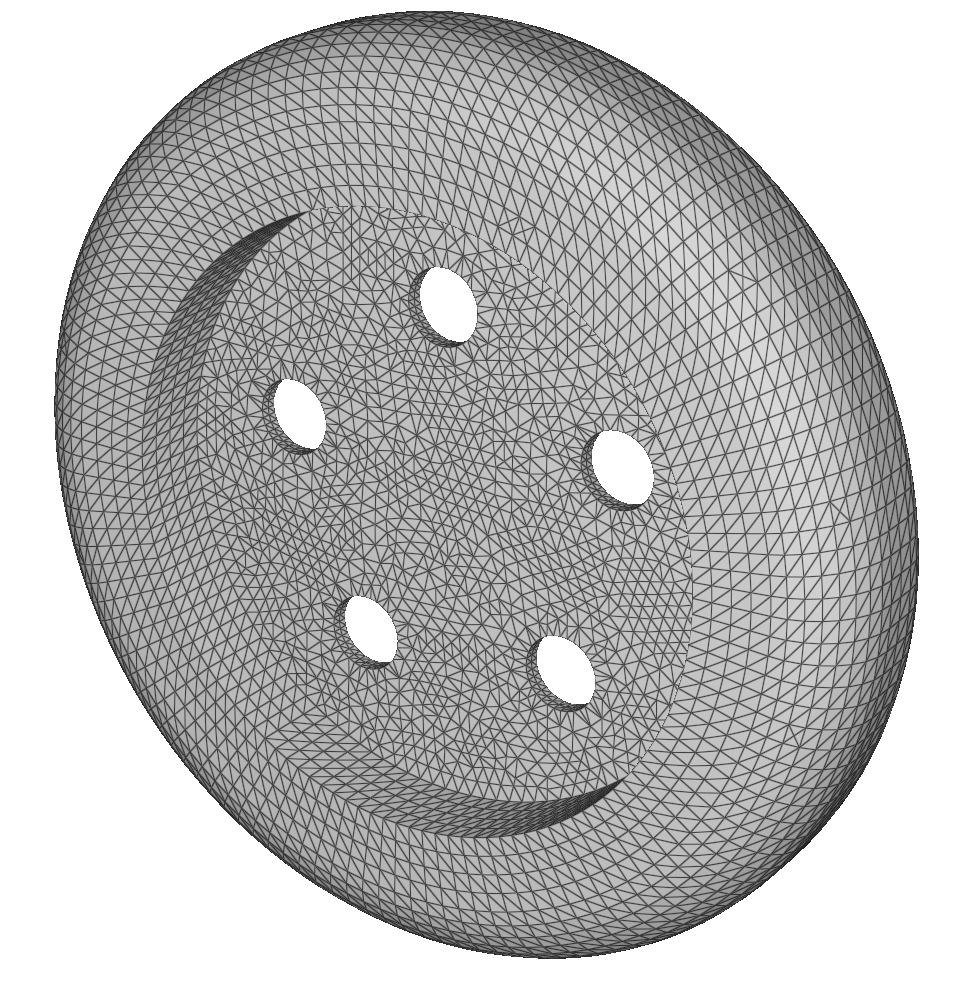}%
    \caption{Tire}
  \end{subfigure}\hfill%
  \begin{subfigure}{0.125\textwidth}%
    \includegraphics[height=2.5cm]{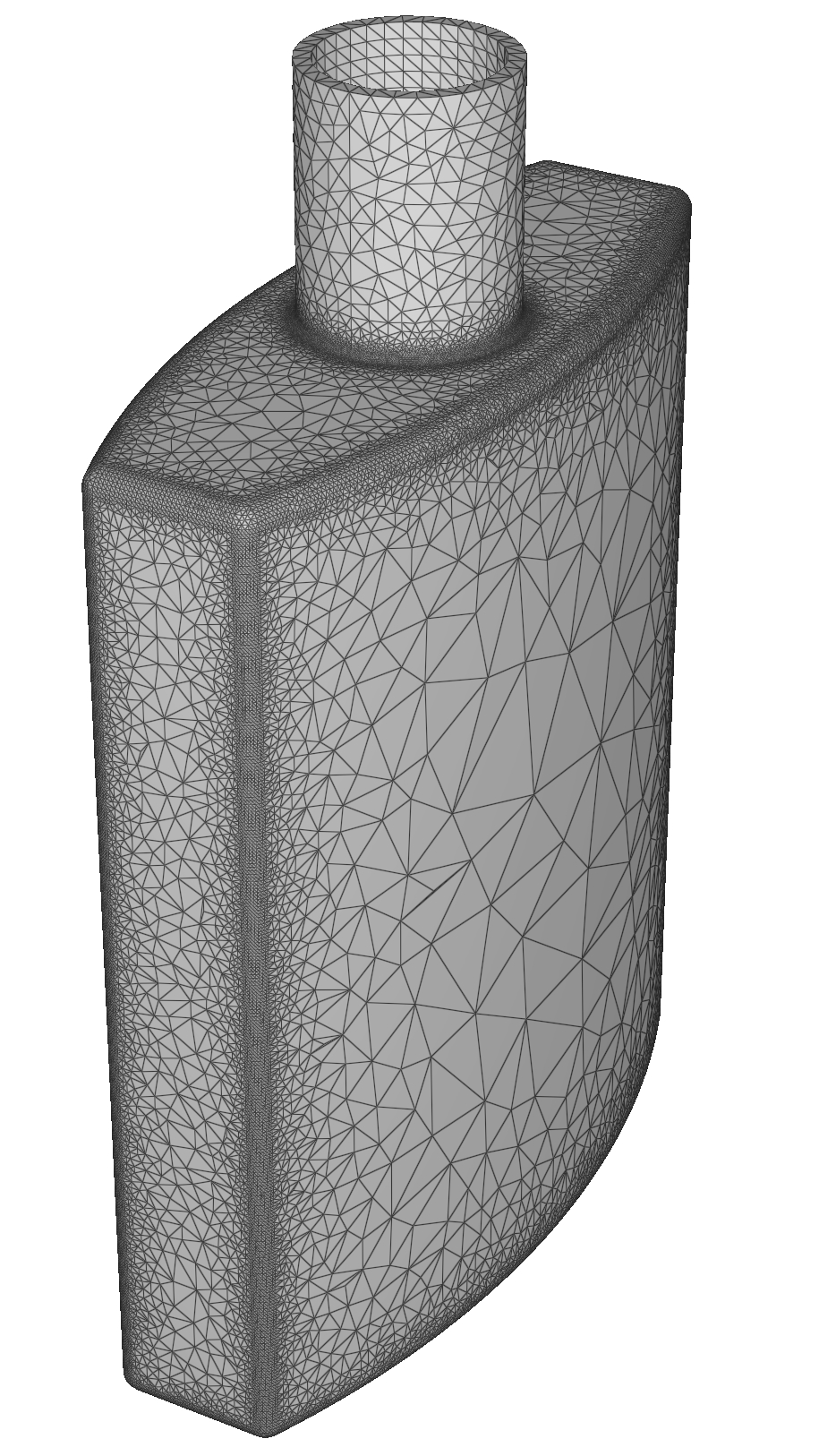}%
    \caption{Bottle}
  \end{subfigure}\hfill%
  \begin{subfigure}{0.1\textwidth}%
    \includegraphics[height=2.5cm]{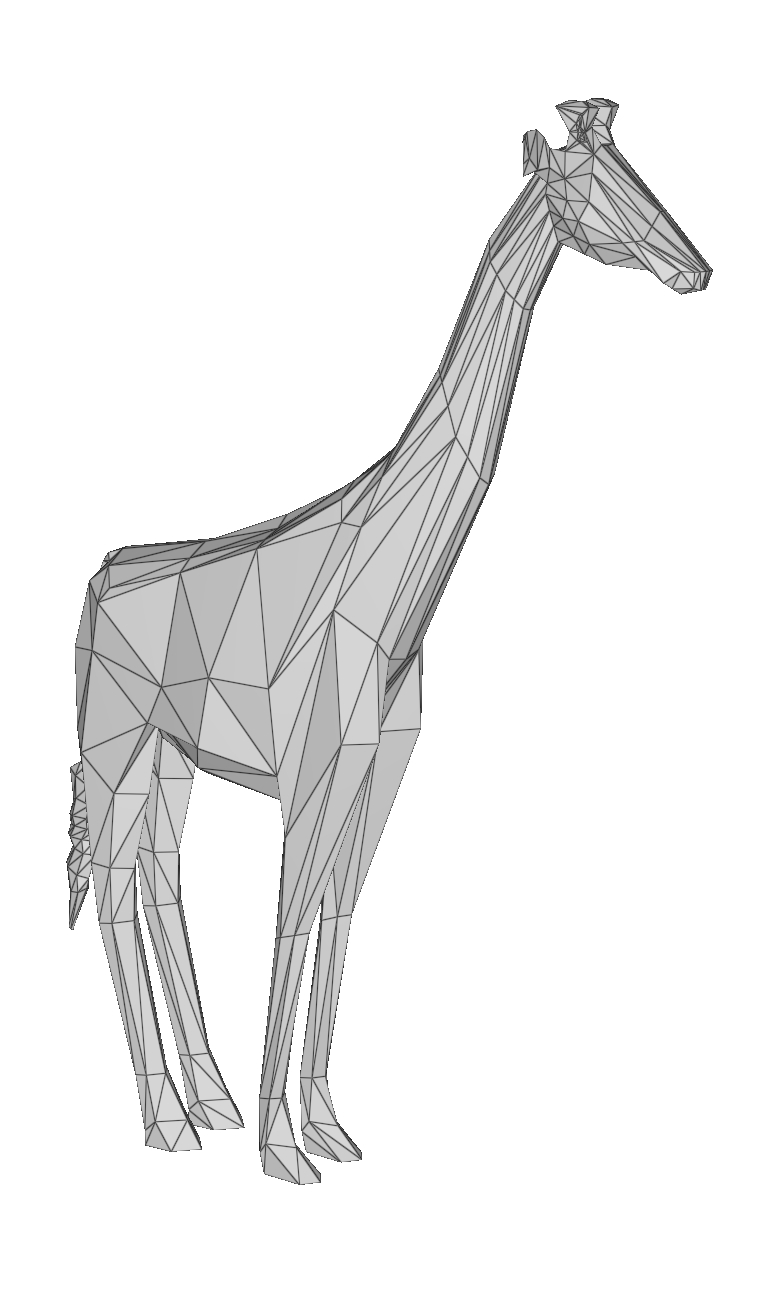}%
    \caption{Giraffe}
  \end{subfigure}%
  \\ \vspace{10pt}%
  \begin{subfigure}{0.425\textwidth}%
    \includegraphics[height=0.3\textwidth]{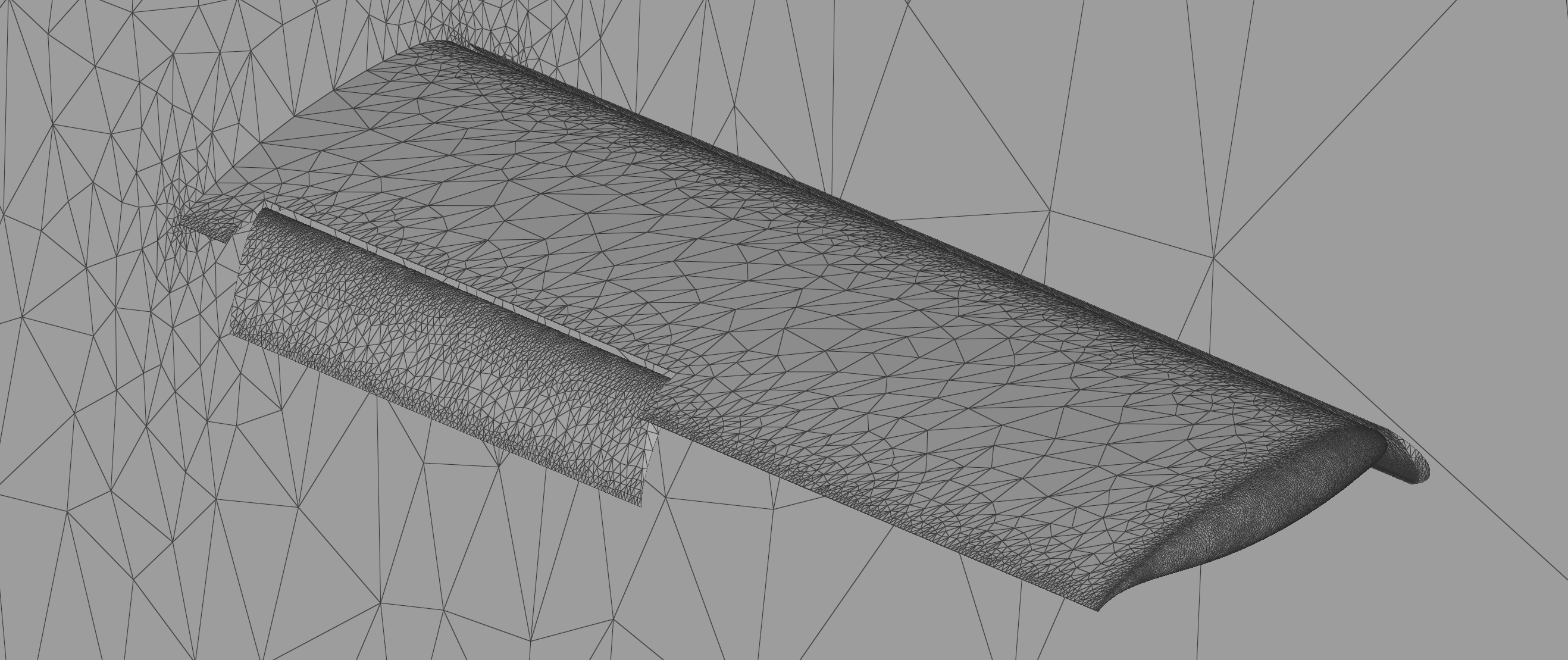}%
    \caption{Simplified High-Lift Common\\Research Model (CRM-HLS)}
  \end{subfigure}%
  \begin{subfigure}{0.475\textwidth}%
    \includegraphics[height=0.32\textwidth]{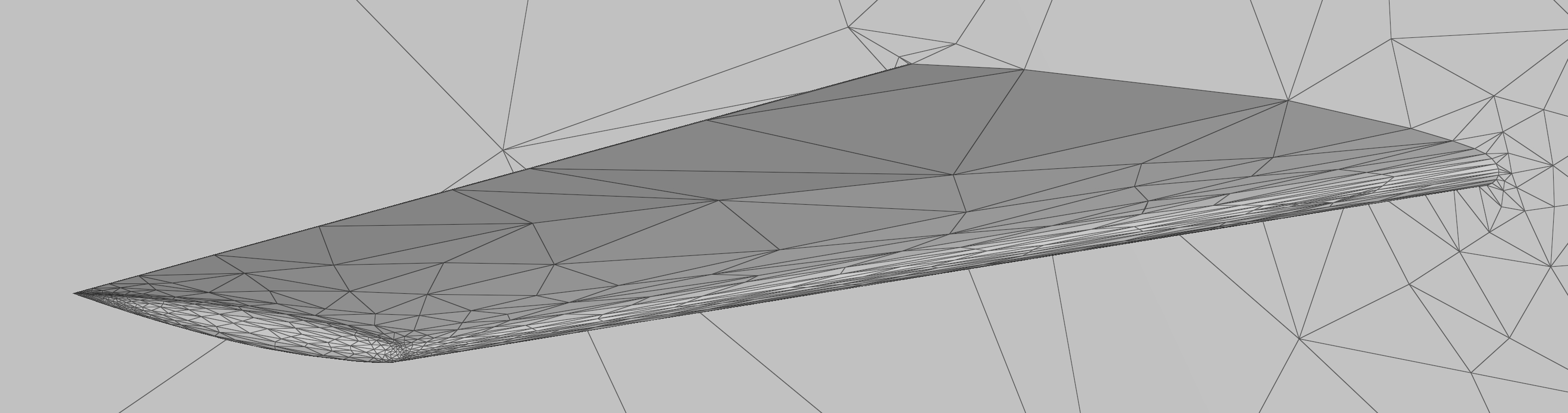}%
    \caption{ONERA-M6}
  \end{subfigure}%
  \\ \vspace{10pt}%
  \begin{subfigure}{0.3\textwidth}%
    \includegraphics[height=0.8\textwidth]{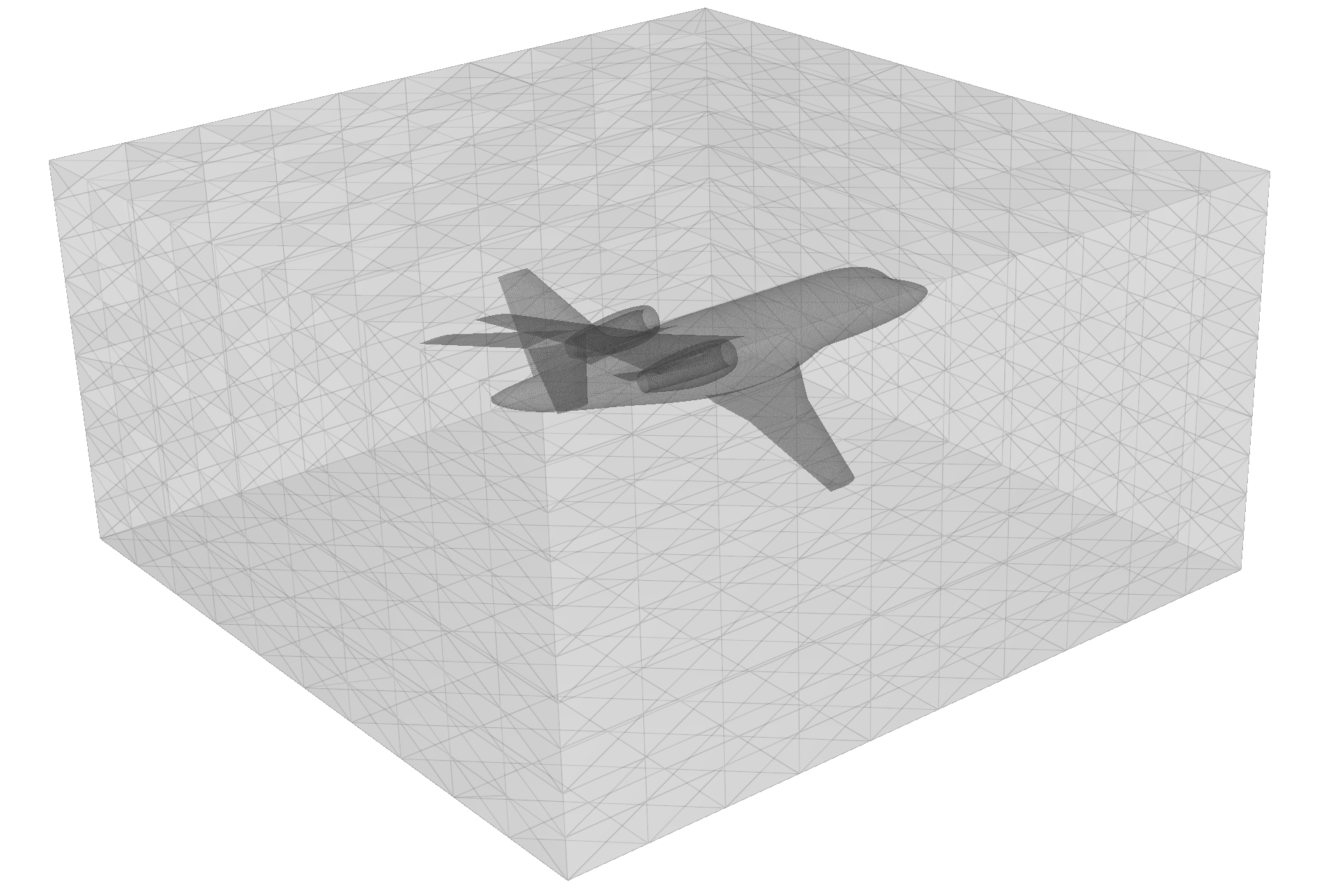}%
    \caption{Falcon}
  \end{subfigure}%
  \begin{subfigure}{0.3\textwidth}%
    \includegraphics[height=0.8\textwidth]{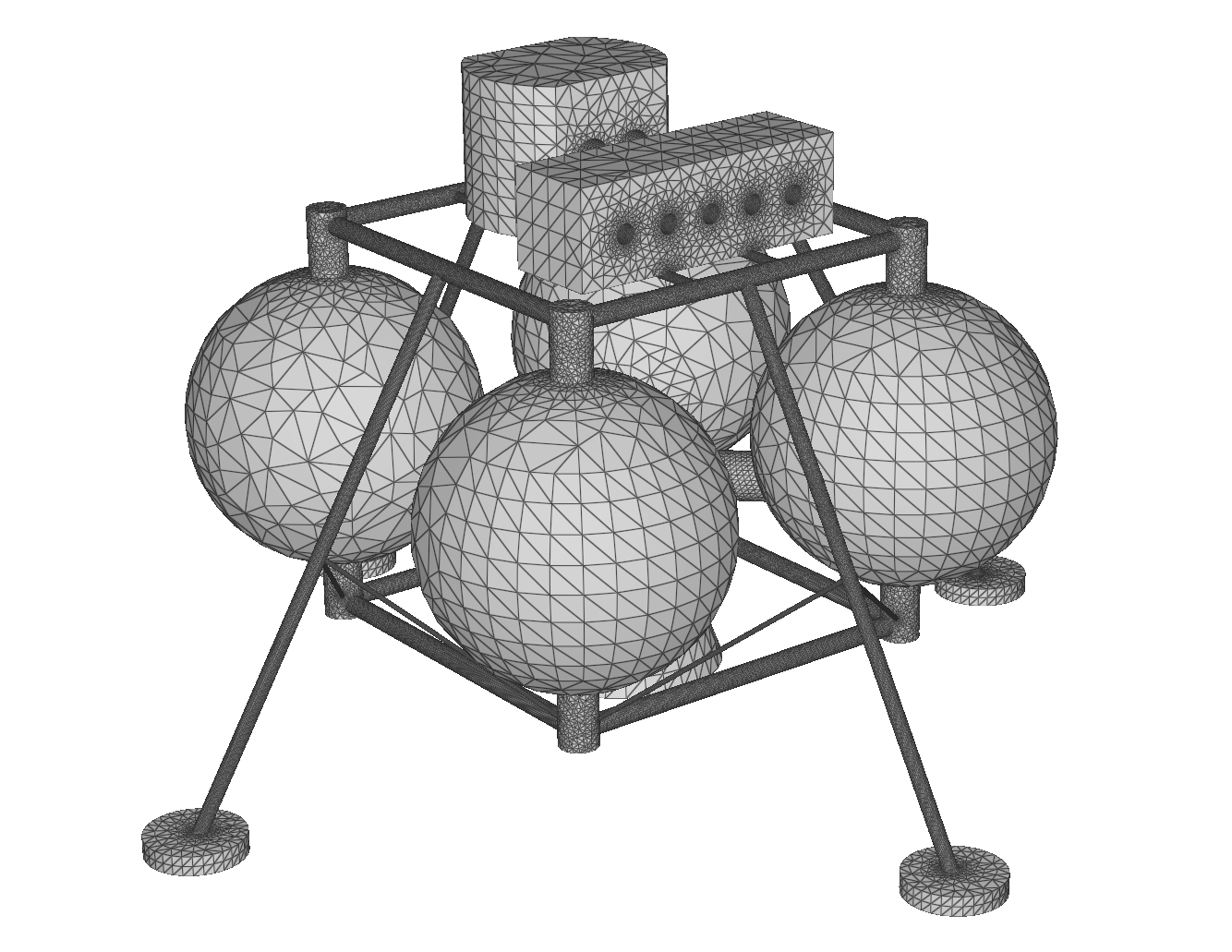}%
    \caption{Lander}
  \end{subfigure}%
  \begin{subfigure}{0.3\textwidth}%
    \includegraphics[height=0.8\textwidth]{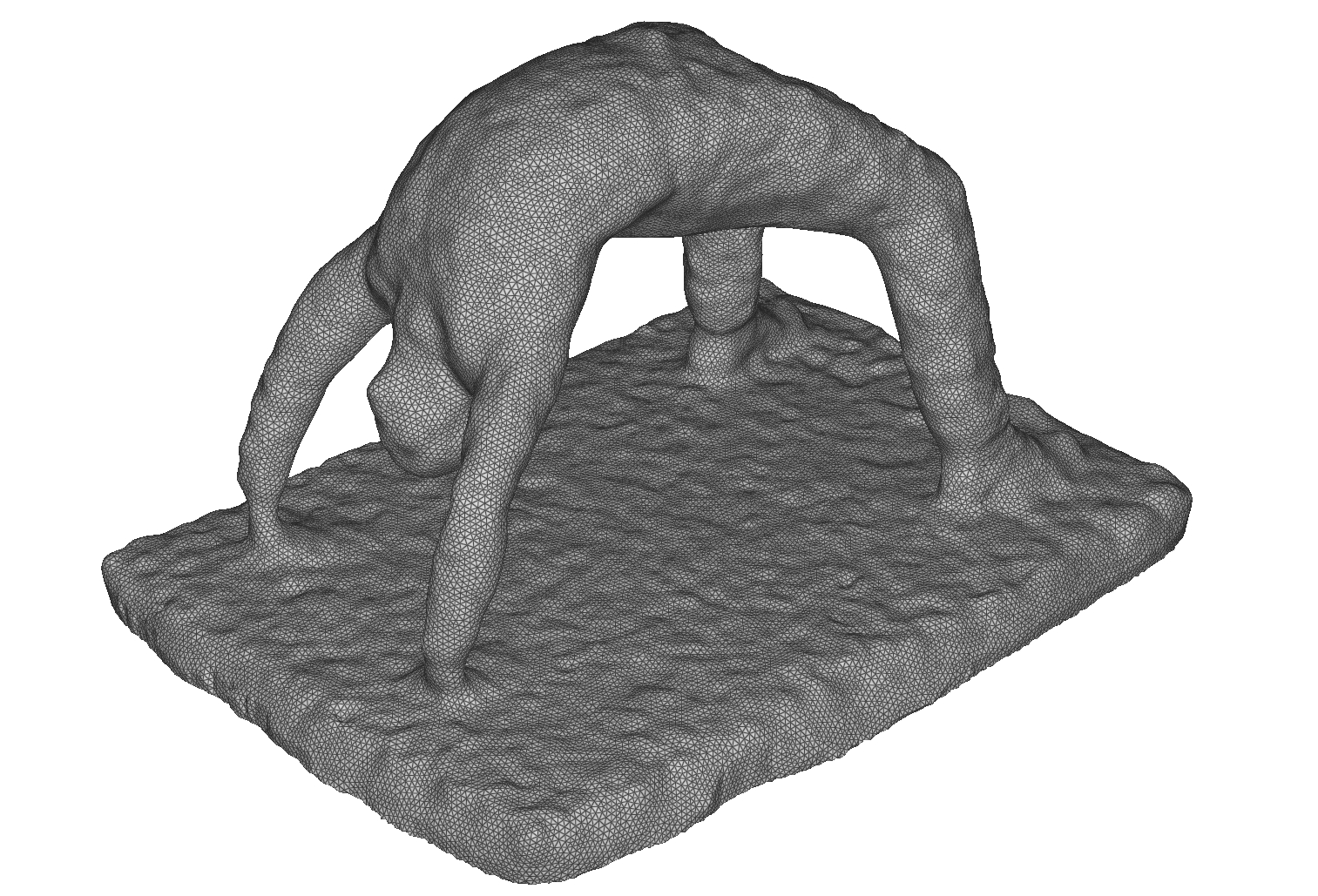}%
    \caption{Pittsburgh Bridge}
  \end{subfigure}%
  \caption{Final recovered surface meshes (possibly with Steiner vertices) for the $3d$ test cases.}
  \label{fig:surfaces-3d}
\end{figure*}

The surface meshes for these test cases, illustrated in \cref{fig:surfaces-3d}, cover a range of geometric complexity and demonstrate the ARO algorithm works very well in $3d$.
\Cref{tab:results-3d} shows the initial number of boundary constraints ($\left|\mathcal{B}_0\right|$) as well as the fraction of the number of these constraints ($\%\,\mathcal{B}_0$) that appear in the initial set of constraints after computing the Delaunay mesh, i.e. $\left|\mathcal{B}^*_0\right|/\left|\mathcal{B}_0\right|$.
The column corresponding to $\%\,\mathcal{B}_f$ shows that the final number of constraints exactly conforms to the representation of the surface, which may be refined through the addition of Steiner vertices.
Most Steiner vertices are computed from intersections with constraint edges ($s_{\diagup}$), though some models required some Steiner vertices on constraint triangles ($s_{\triangle}$).
The time to compute the Delaunay triangulation (DT) and to run the ARO algorithm is very reasonable -- the rest of the time reported in the ``Total" column also includes the time to insert Steiner vertices, which can be quite costly.
\begin{table*}[h!]
  \centering
  \resizebox{\textwidth}{!}{
  \begin{tabular}{rlllllllllll}
     & $\left|\mathcal{{V}}_0\right|$ & $\left|\mathcal{{B}}_0\right|$ & $\%\,\mathcal{{B}}_0$ & $\left|\mathcal{{B}}_f\right|$ & $\%\,\mathcal{{B}}_f$ & $\left|\mathcal{M}\right|$ & DT (s) & {{ARO}} (s) & Total (s) & $s_{\diagup}$ & $s_{\triangle}$ \\ \hline
Bunny & 5,884 & 11,764 & 99.753 \% & 11,764 & 100.0 \% & 18,765 & 0.034 & 0.0 & 0.05 & 0 & 0\\ 
Spot & 2,930 & 5,856 & 97.575 \% & 5,856 & 100.0 \% & 9,918 & 0.017 & 0.001 & 0.02 & 0 & 0\\ 
Pool Toy & 3,087 & 6,174 & 99.627 \% & 6,174 & 100.0 \% & 12,773 & 0.018 & 0.0 & 0.03 & 0 & 0\\ 
Tire & 7,450 & 14,916 & 78.667 \% & 14,950 & 100.0 \% & 24,621 & 0.165 & 0.051 & 0.28 & 17 & 0\\ 
Bottle & 29,709 & 59,414 & 89.582 \% & 59,648 & 100.0 \% & 117,997 & 0.437 & 0.103 & 0.98 & 116 & 1\\ 
Giraffe & 471 & 938 & 49.787 \% & 1,080 & 100.0 \% & 1,622 & 0.002 & 0.006 & 0.02 & 50 & 21\\ 
CRM HLS & 169,187 & 338,386 & 97.286 \% & 338,406 & 100.0 \% & 719,668 & 1.671 & 0.064 & 3.37 & 8 & 2\\ 
ONERA-M6 & 837 & 1,670 & 69.76 \% & 1,700 & 100.0 \% & 2,736 & 0.004 & 0.005 & 0.02 & 14 & 1\\ 
Falcon & 76,738 & 153,476 & 99.243 \% & 153,504 & 100.0 \% & 523,465 & 0.849 & 0.022 & 1.39 & 14 & 0\\ 
Lander & 233,782 & 467,668 & 95.691 \% & 467,966 & 100.0 \% & 1,451,198 & 3.756 & 0.29 & 9.93 & 149 & 0\\ 
Pittsburgh Bridge & 75,081 & 150,170 & 99.344 \% & 150,170 & 100.0 \% & 242,806 & 0.884 & 0.005 & 1.1 & 0 & 0\\ 

  \end{tabular}}
  \caption{Initial ($\left|\mathcal{B}_0\right|$) and final ($\left|\mathcal{B}_f\right|$) number of boundary triangles after the boundary recovery procedure.
    $\%\,\mathcal{B}_0$ refers to the fraction of $\mathcal{B}_0$ present in the initial Delaunay mesh.
    The final meshes completely conform to the boundary representation (possibly with Steiner vertices) since $\%\,\mathcal{B}_f$ is $100\%$ for all test cases.
    The final number of tetrahedra is $\left|\mathcal{M}\right|$; $s_{\diagup}$ and $s_{\triangle}$ refer to the number of Steiner vertices added to the constraint edges and triangles, respectively.
    The time for the initial Delaunay triangulation (DT) of the input vertices $\mathcal{V}_0$ and for the ARO algorithm are also reported in seconds.
  }
  \label{tab:results-3d}
\end{table*}
\subsection{Boundary-conforming~pentatopizations.\hfill}
The ability of the ARO algorithm to recover $4d$ tetrahedralizations will now be assessed.
All tetrahedralizations below were generated by extruding an initial surface triangulation (some of which were introduced in the previous section) into $4d$ by subdividing the resulting prisms.
All cases involve some form of motion, whether it be a uniform scaling or rotation.
The extrusion procedure occurs over a prescribed number of time steps, where the time step is taken to be the minimum edge length in the surface mesh.
The vertices on the surface are then transformed by the prescribed motion while keeping the boundary triangle connectivity fixed.
This results in tetrahedralizations along the lateral walls of the spacetime domain.
To form a closed tetrahedralization, the tetrahedralizations at the initial and final time steps are added to the boundary mesh.
The latter meshes are computed using \texttt{TetGen} in order to create quality meshes at the extremities of the domain.

\Cref{tab:results-4d} summarizes the results, with each test case elaborated upon below.

\begin{figure*}[h!]
  \centering
  \includegraphics[width=0.3\textwidth]{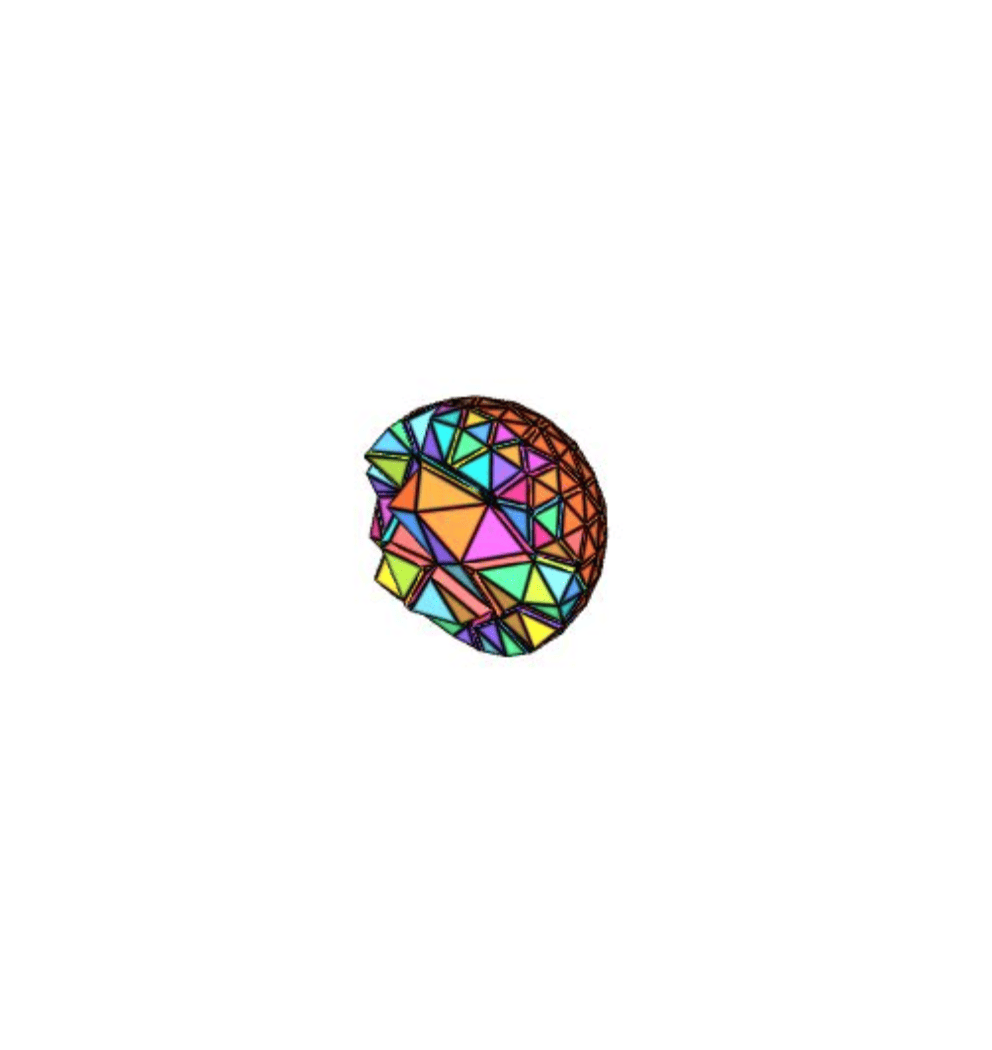}
  \includegraphics[width=0.3\textwidth]{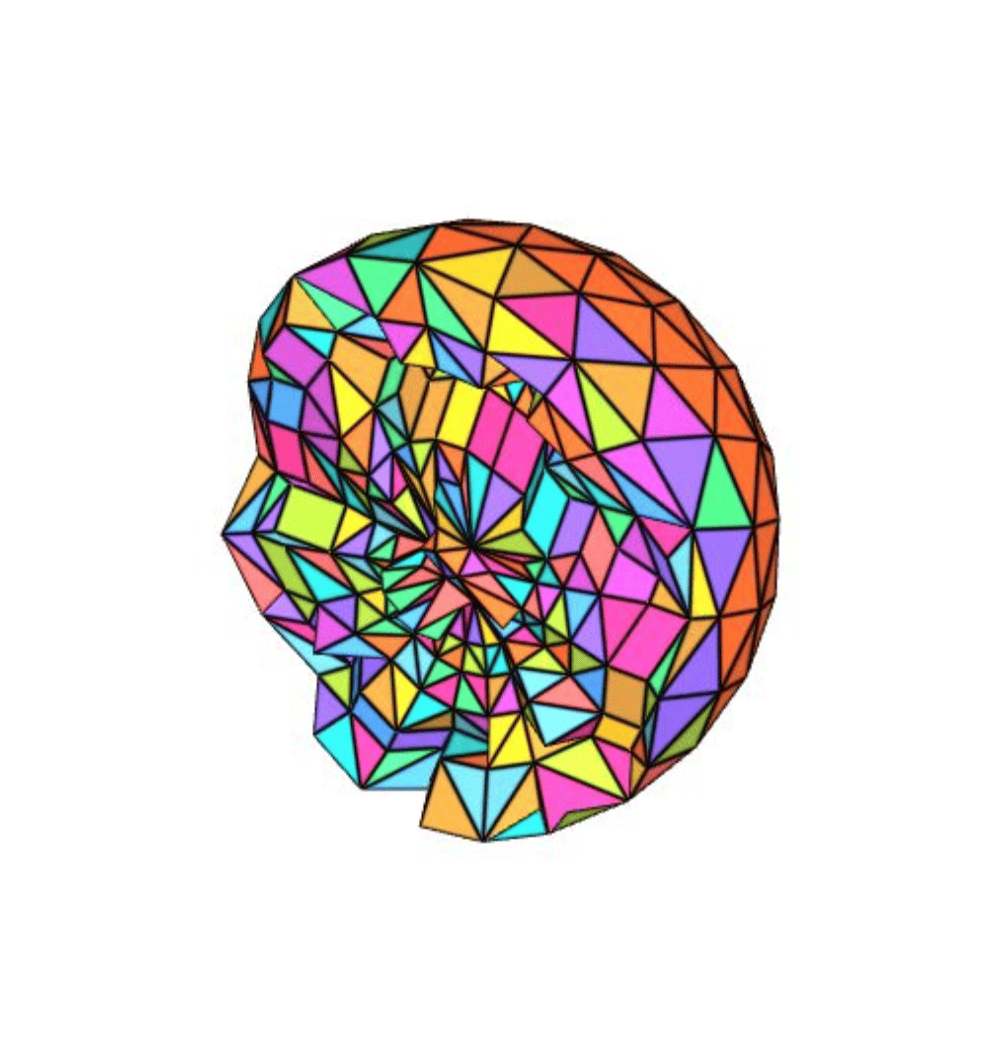}
  \includegraphics[width=0.3\textwidth]{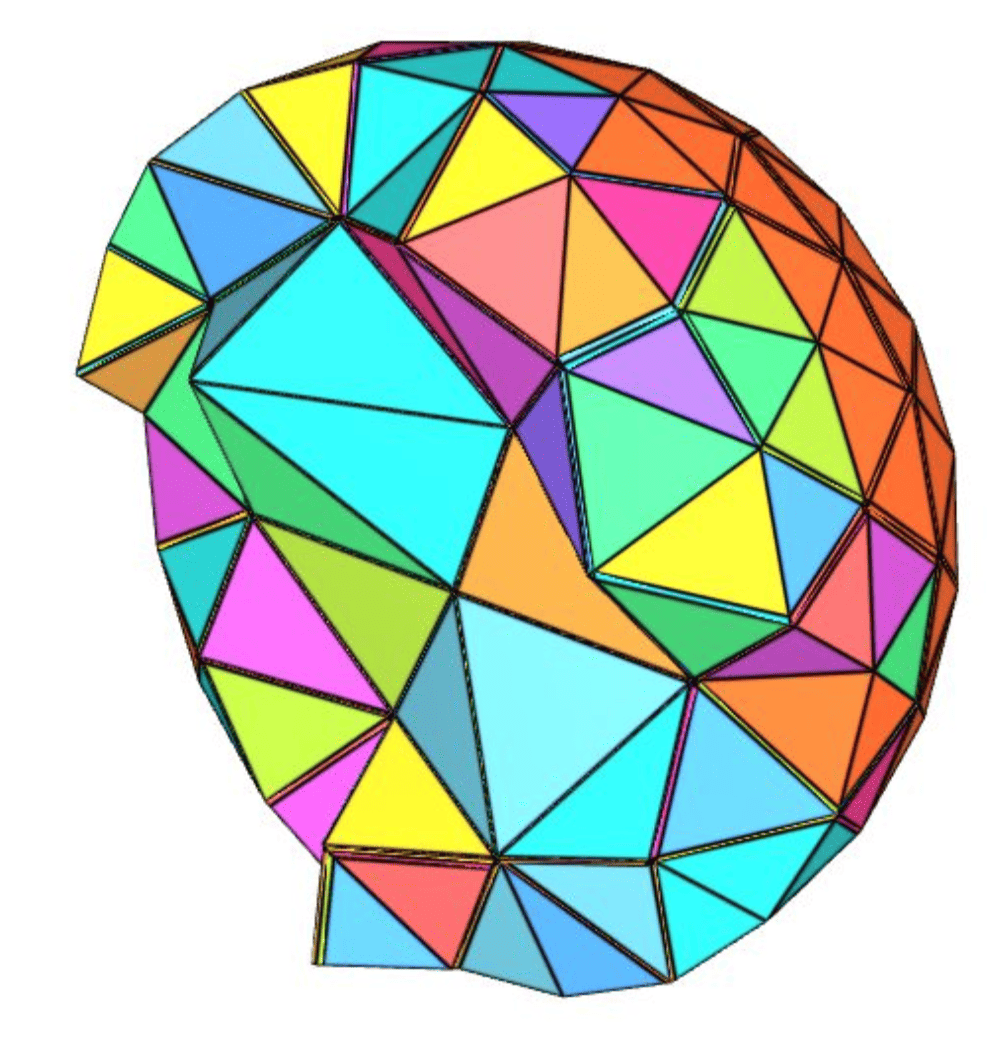}
  \caption{Slices of the resulting pentatopizations for the expanding and rotating sphere at $t = 0$ (left), $t = 1.38$ (middle) and $t = 2.76$ (right) seconds.}
  \label{fig:spacetime-sphere}
\end{figure*}

\paragraph{Expanding and Rotating Sphere.} The first case consists of an expanding and rotating sphere, which starts with a unit radius.
The surface mesh is a uniform subdivision of an icosahedron with 162 vertices and 320 triangles.
This surface mesh is extruded into $4d$ with a total of 10 time slabs (11 temporal slices), thus producing a boundary tetrahedralization with 10,995 tetrahedra.
At each time $t$, a point on the surface is first rotated about the $x$-axis with an angle of $\theta(t) = 0.1t$ and then expanded by a factor of $(1 + t)$.
\Cref{fig:spacetime-sphere} shows that the expansion is quite large relative to the initial radius.
The initial Delaunay mesh only contains about 32\% of the boundary tetrahedra (7425 tetrahedral constraints were missing).
The first iteration of the ARO algorithm was able to recover almost all of the constraints with only 78 constraints missing.
After adding a total of 88 Steiner vertices and restarting the ARO algorithm, a complete boundary-conforming mesh of the expanding sphere was obtained.
\begin{table*}[h!]
  \centering
  \resizebox{\textwidth}{!}{
  \begin{tabular}{rllllllllllll}
     & $\left|\mathcal{{V}}_0\right|$ & $\left|\mathcal{{B}}_0\right|$ & $\%\,\mathcal{{B}}_0$ & $\left|\mathcal{{B}}_f\right|$ & $\%\,\mathcal{{B}}_f$ & $\left|\mathcal{M}\right|$ & DT (s) & {{ARO}} (s) & Total (s) & $s_{\diagup}$ & $s_{\triangle}$ & $s_{\tetrahedron}$\\ \hline
Sphere & 1,872 & 10,995 & 32.469 \% & 11,381 & 100.0 \% & 16,151 & 0.881 & 0.324 & 1.45 & 29 & 58 & 1\\ 
Puck & 3,998 & 23,675 & 31.958 \% & 24,397 & 100.0 \% & 41,907 & 0.169 & 0.628 & 2.18 & 75 & 78 & 0\\ 
Pool Toy & 10,230 & 61,224 & 71.431 \% & 61,757 & 100.0 \% & 94,025 & 0.469 & 0.867 & 16.7 & 73 & 39 & 0\\ 
Bunny & 19,872 & 118,424 & 71.645 \% & 118,973 & 99.991 \% & 255,967 & 0.991 & 1.593 & 53.07 & 95 & 24 & 0\\ 
Spot & 10,494 & 63,244 & 70.677 \% & 63,519 & 99.997 \% & 135,766 & 0.504 & 0.798 & 11.24 & 34 & 25 & 0\\ 
CRM-HLS & 1,427,532 & 8,592,853 & 58.285 \% & 8,592,853 & 99.255 \% & 26,021,888 & 220.506 & 500.407 & 772.13 & 0 & 0 & 0\\ 

  \end{tabular}}
    \caption{Initial ($\left|\mathcal{B}_0\right|$) and final ($\left|\mathcal{B}_f\right|$) number of boundary tetrahedra after the boundary recovery procedure.
    $\%\,\mathcal{B}_0$ refers to the fraction of $\mathcal{B}_0$ present in the initial Delaunay mesh.
    Some of the final meshes completely conform to the boundary representation (possibly with Steiner vertices), but some have $\%\,\mathcal{B}_f$ which is not equal $100\%$.
    The final number of pentatopes is $\left|\mathcal{M}\right|$; $s_{\diagup}$, $s_{\triangle}$ and $s_{\tetrahedron}$ refer to the number of Steiner vertices added to the constraint edges, triangles and tetrahedra, respectively.
    The time for the initial Delaunay triangulation (DT) of the input vertices $\mathcal{V}_0$ and for the ARO algorithm are also reported in seconds.
  }
  \label{tab:results-4d}
\end{table*}
\begin{figure*}[h!]
  \centering
  \includegraphics[width=0.3\textwidth]{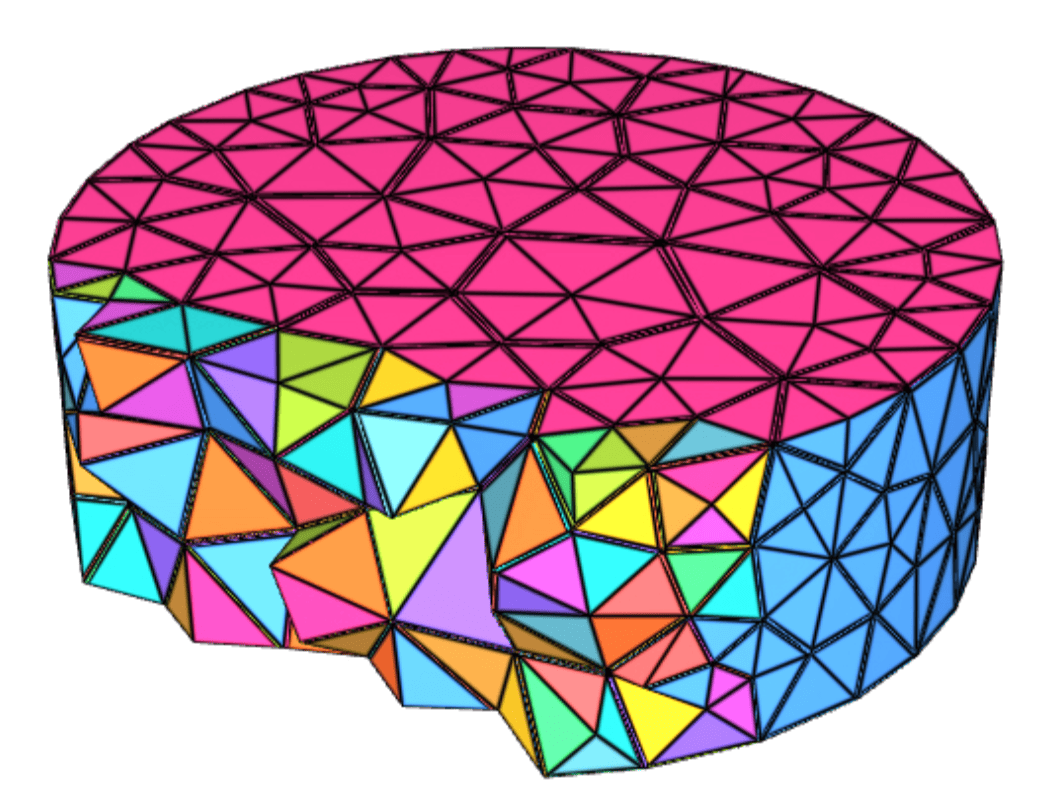}
  \includegraphics[width=0.3\textwidth]{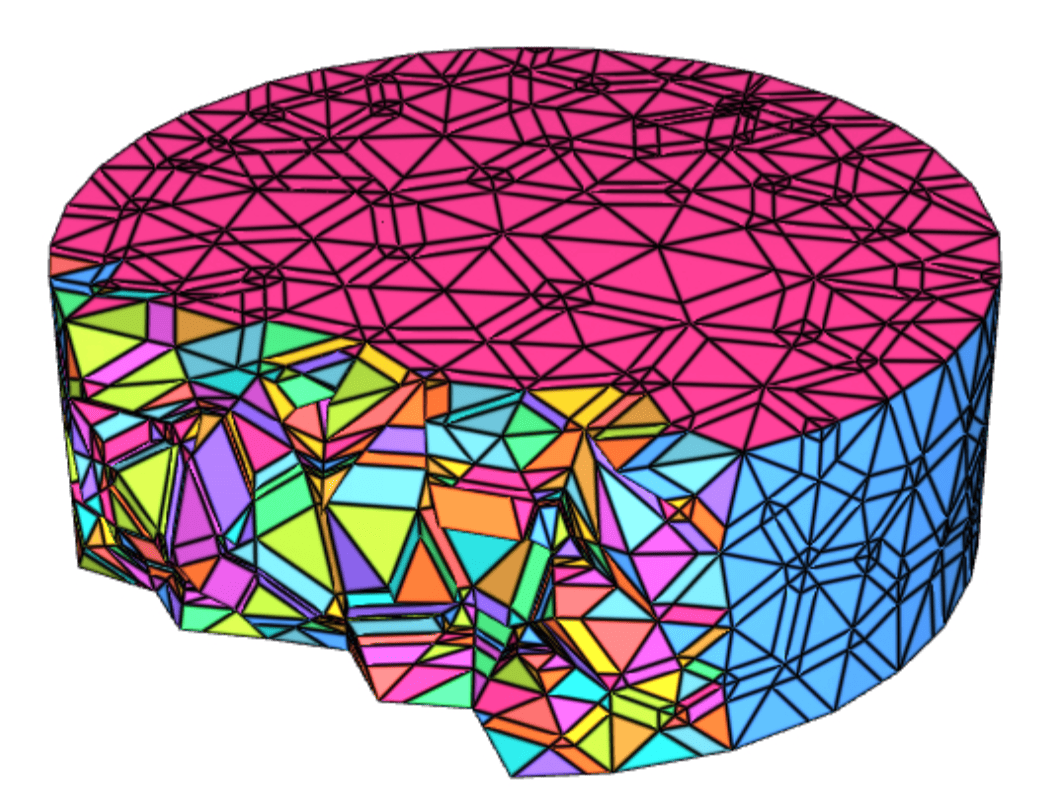}
  \includegraphics[width=0.3\textwidth]{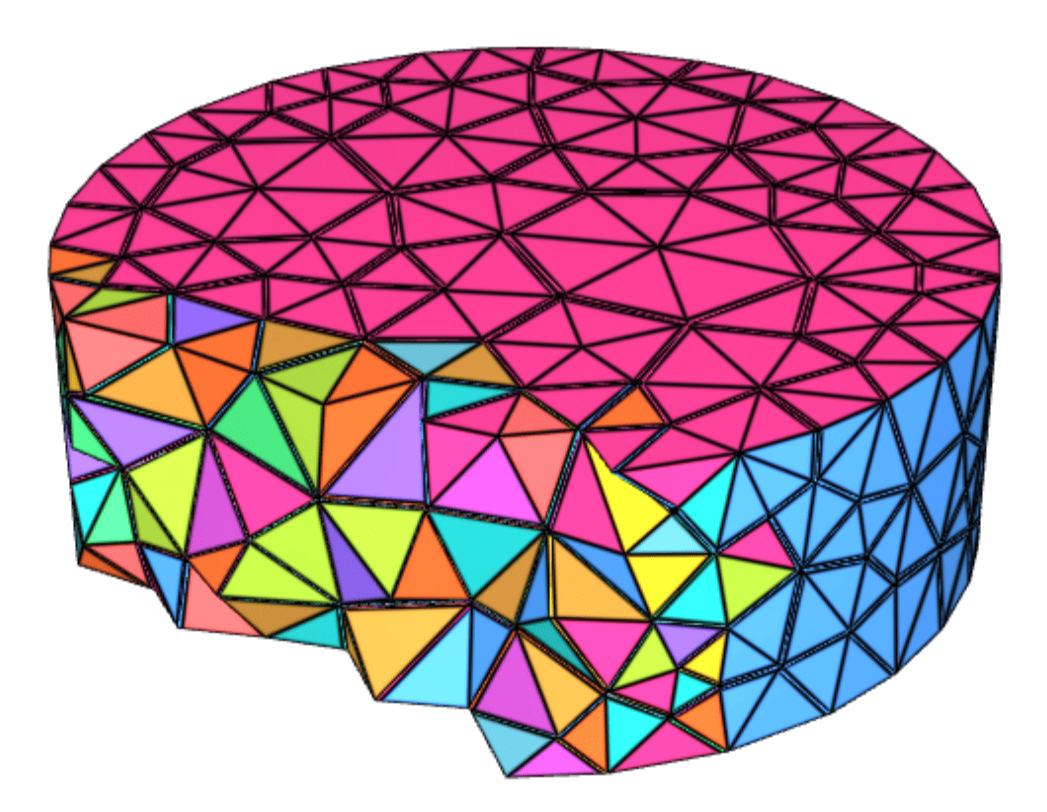}
  \caption{Slices of the resulting pentatopizations for the rotating puck at $t = 0$ (left), $t = 0.368$ (middle) and $t = 0.736$ (right) seconds.}
  \label{fig:spacetime-puck}
\end{figure*}
\begin{figure*}[h!]
  \centering
  \includegraphics[width=0.3\textwidth]{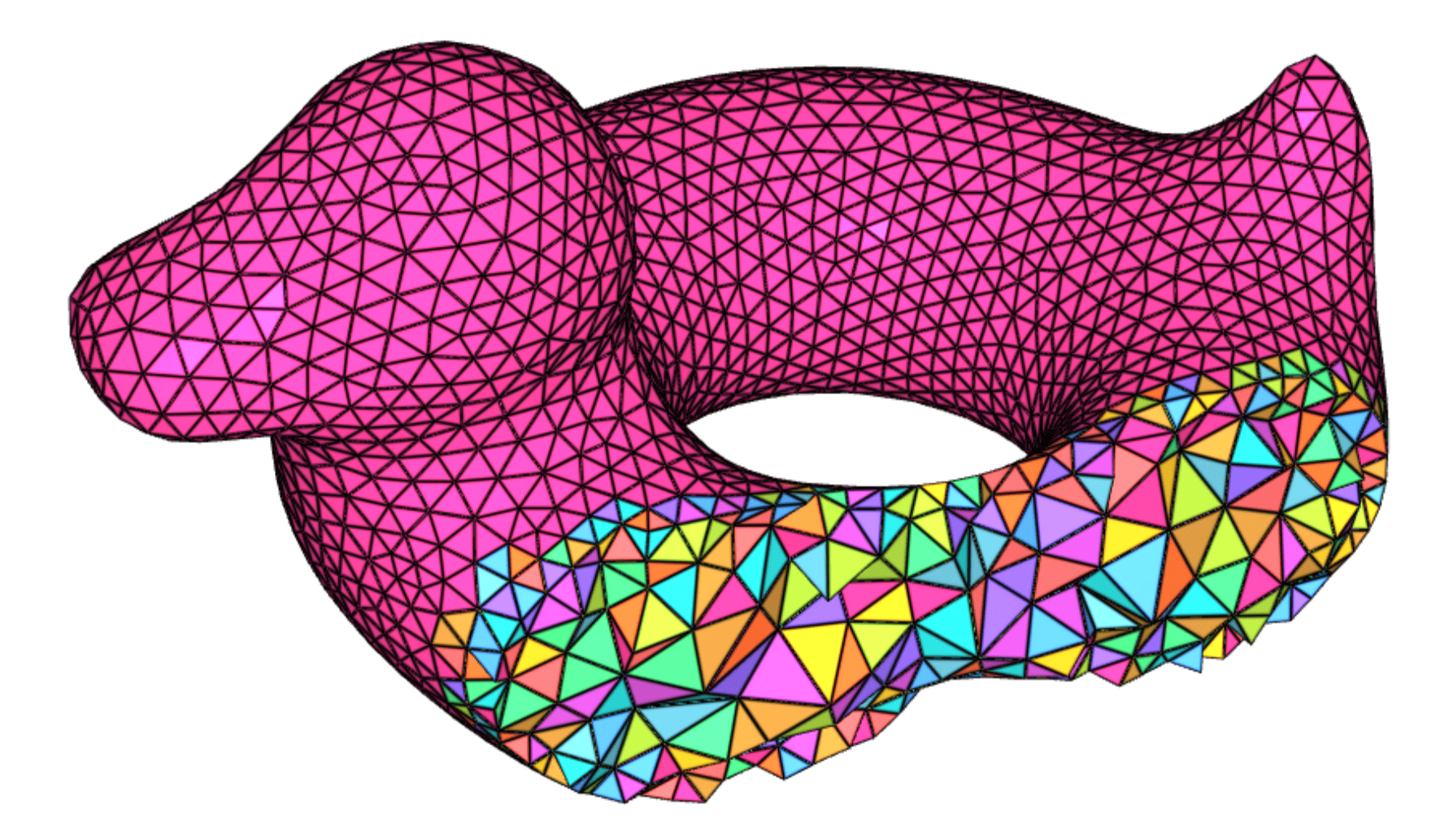}
  \includegraphics[width=0.3\textwidth]{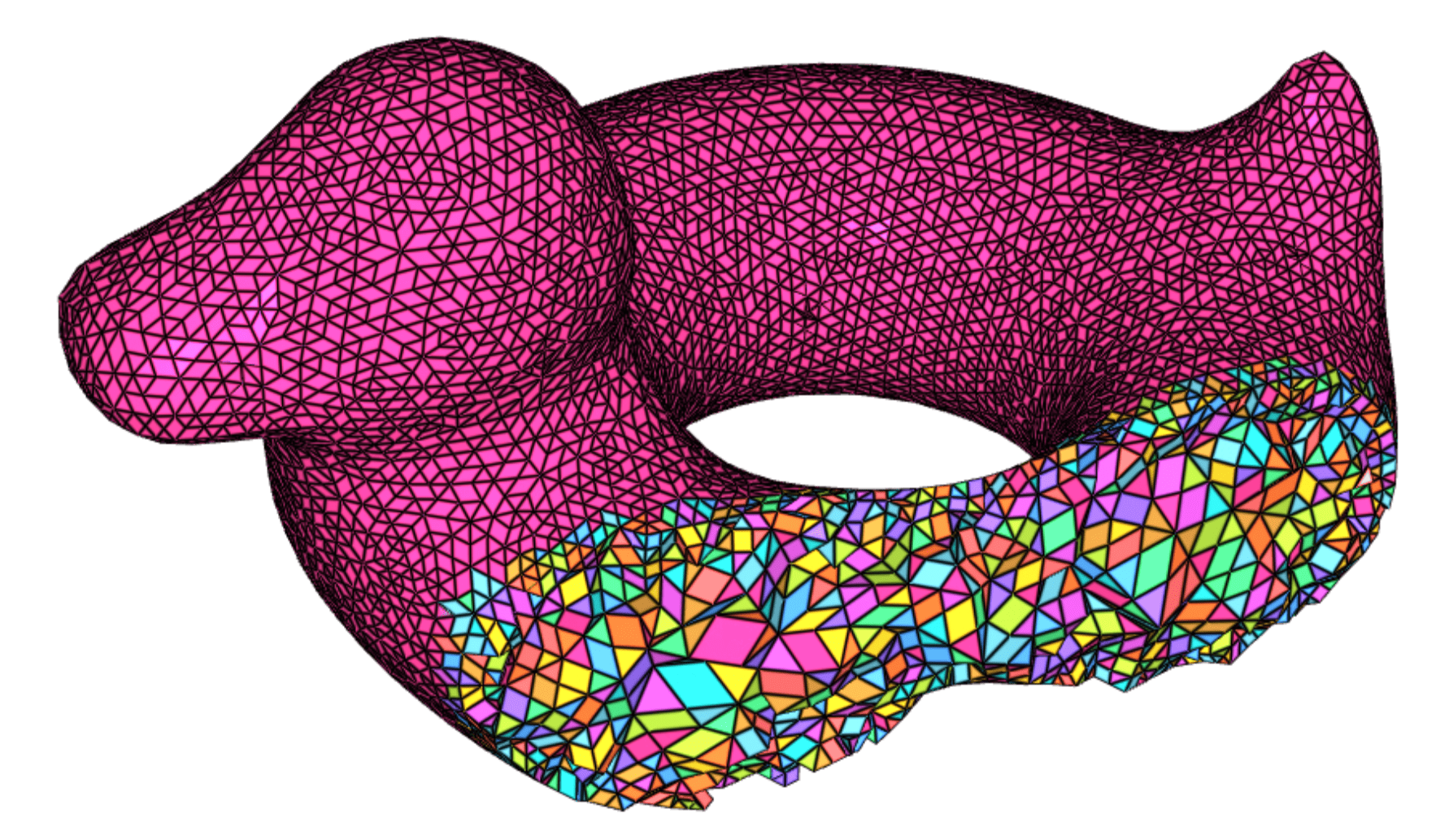}
  \includegraphics[width=0.3\textwidth]{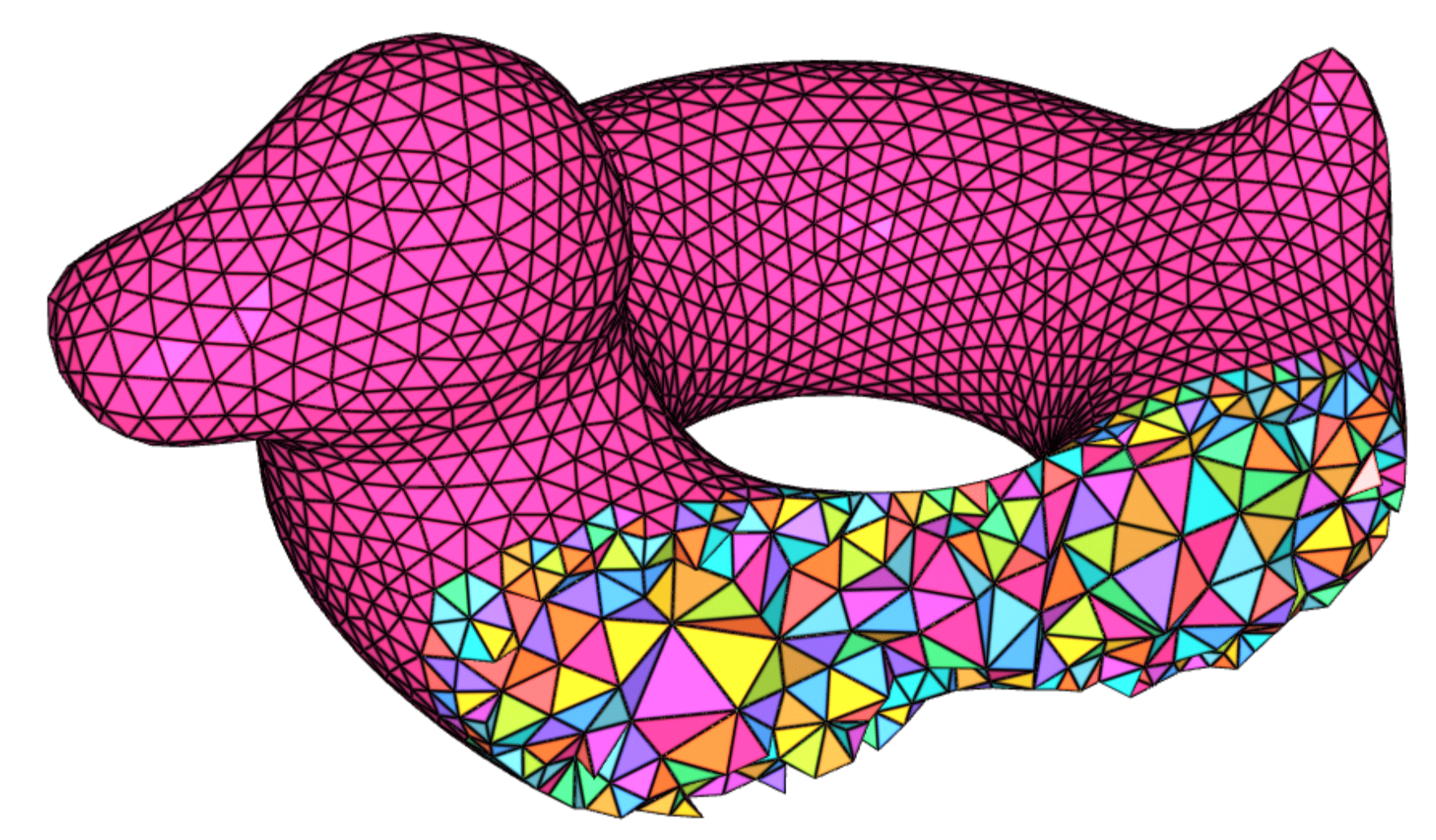}
  \caption{Slices of the resulting pentatopizations for the inflating pool toy at $t = 0$ (left), $t = 0.00151$, (middle) and $t = 0.0302$ (right) seconds.}
  \label{fig:spacetime-duck}
\end{figure*}
\begin{figure*}[!h]
  \centering
  \includegraphics[width=0.3\textwidth]{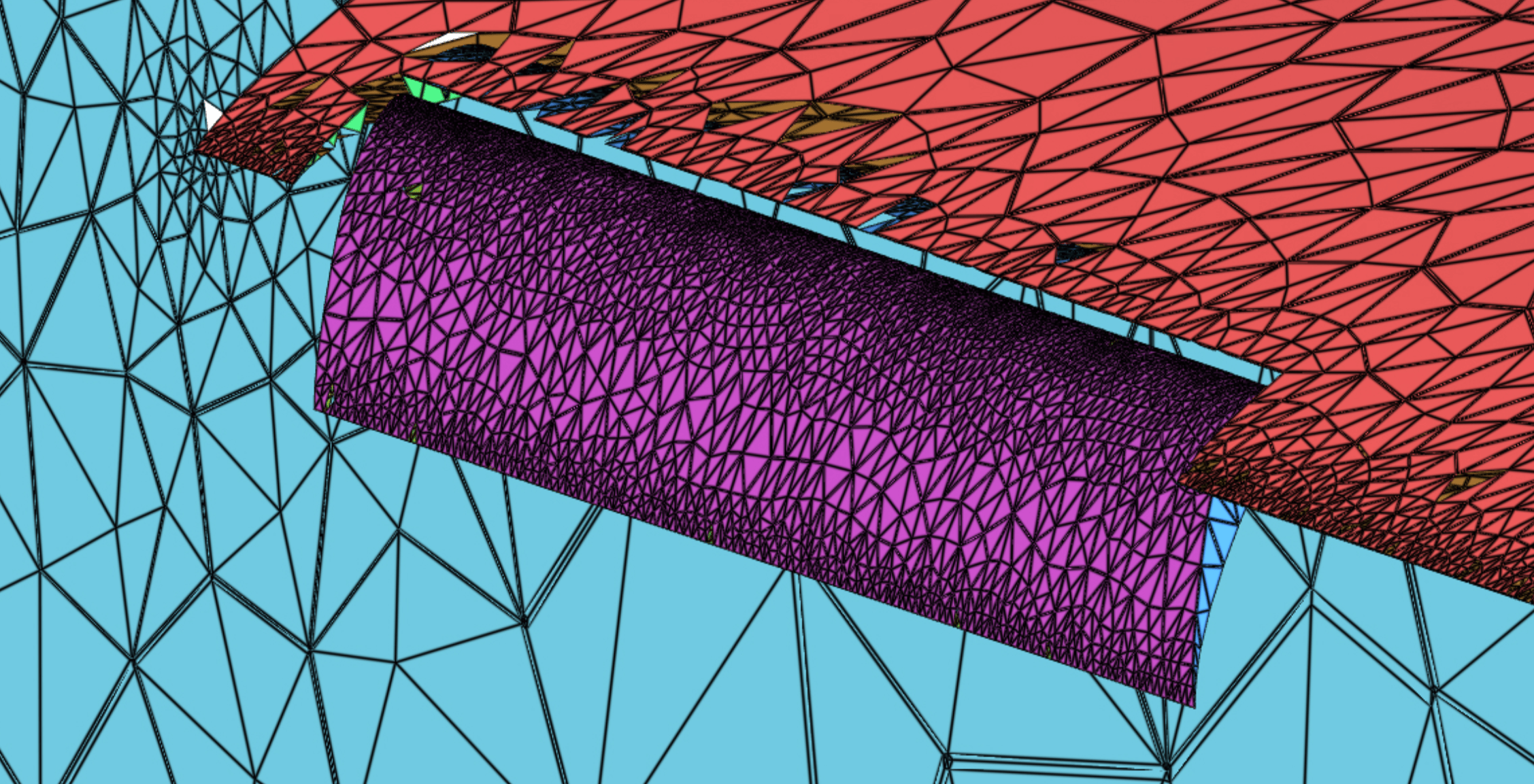}
  \includegraphics[width=0.3\textwidth]{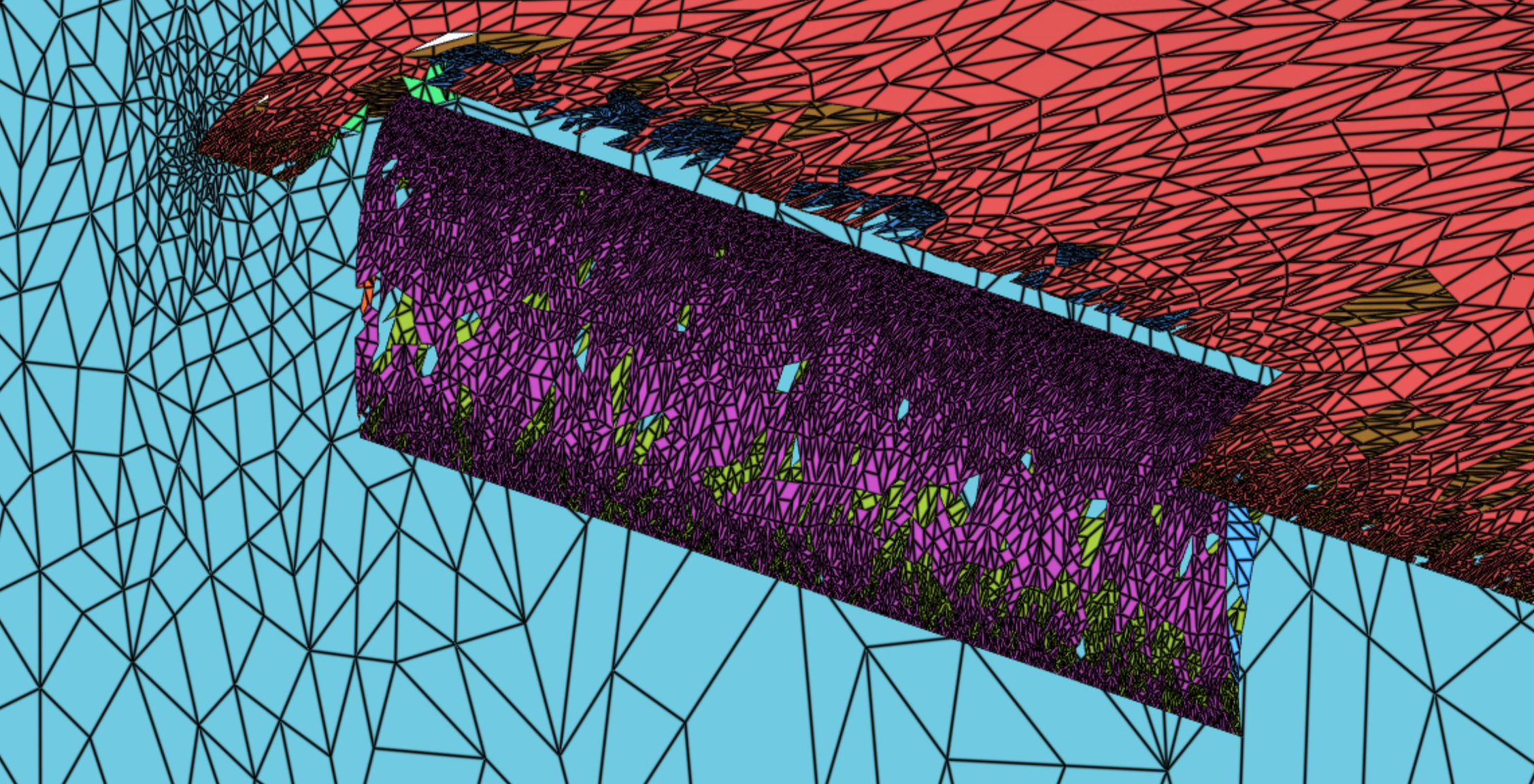}
  \includegraphics[width=0.3\textwidth]{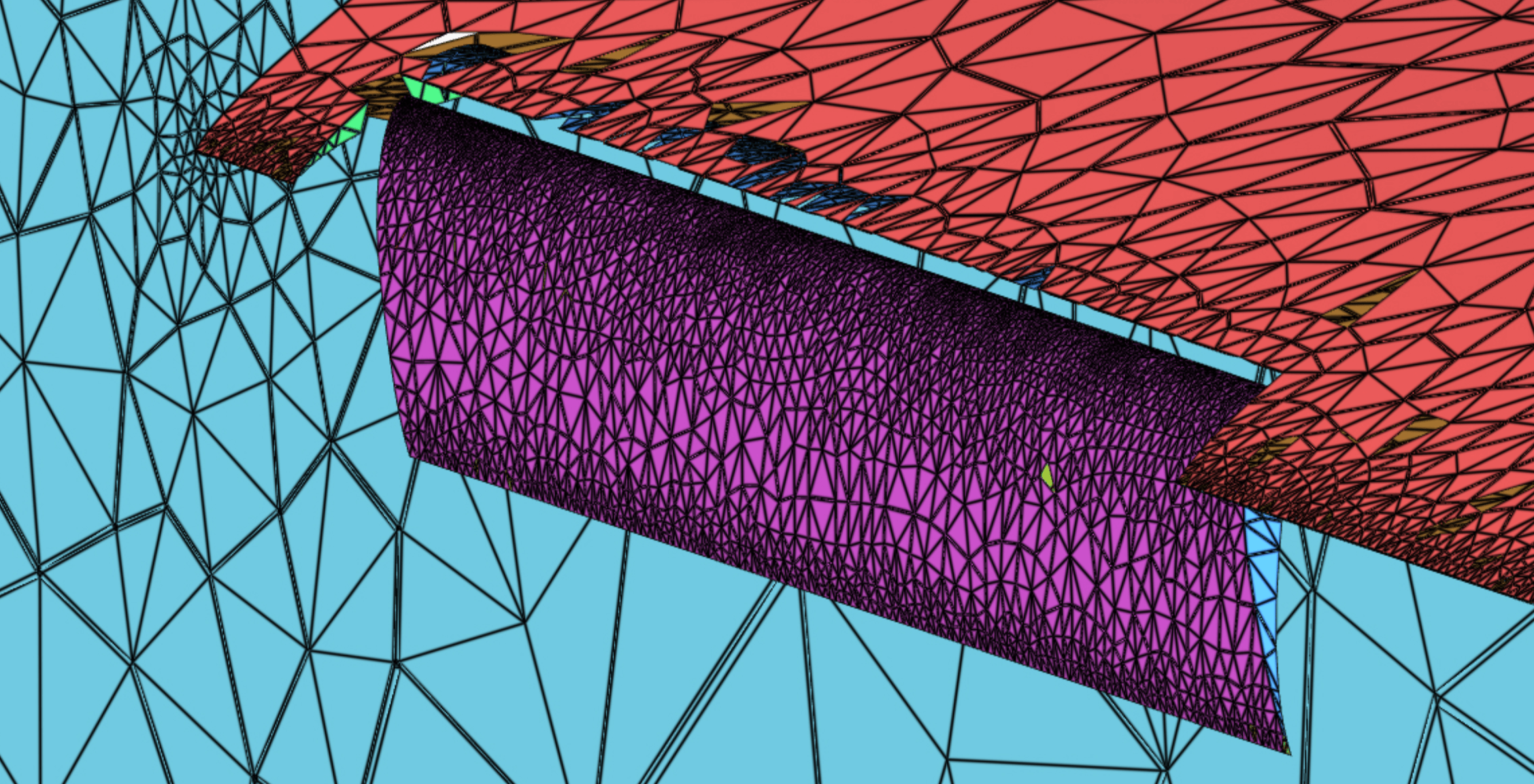}\\
  \includegraphics[width=0.3\textwidth]{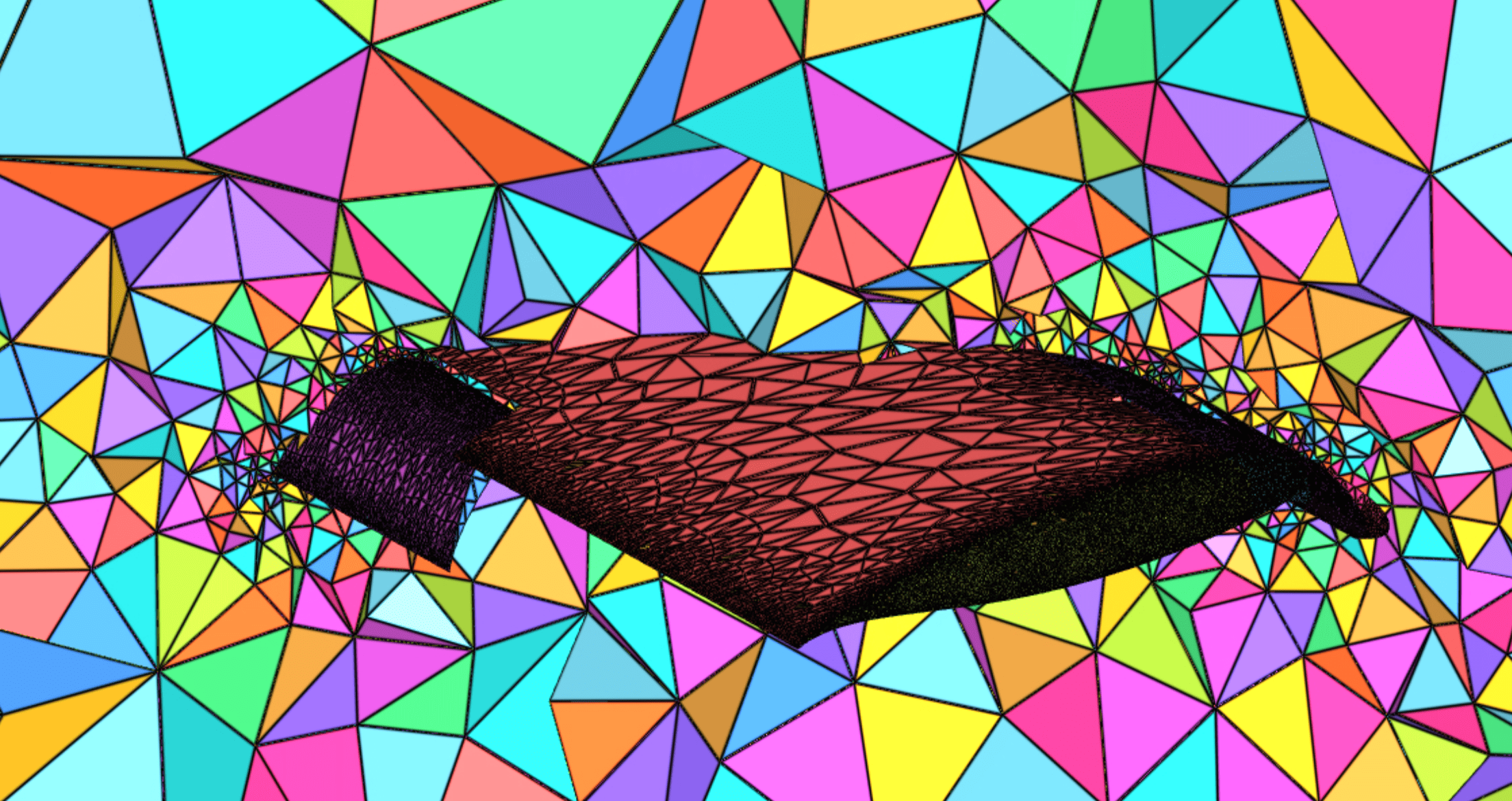}
  \includegraphics[width=0.3\textwidth]{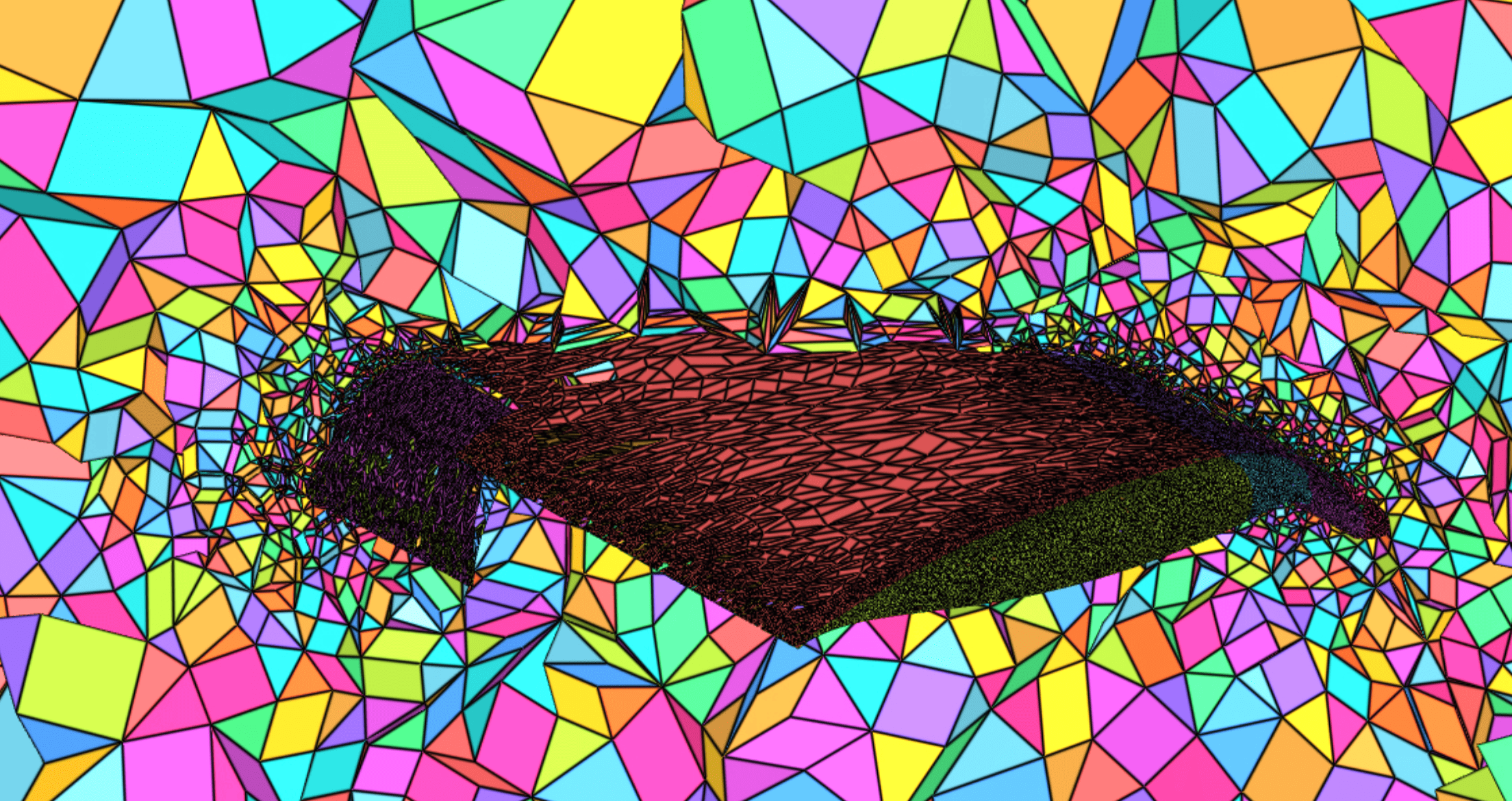}
  \includegraphics[width=0.3\textwidth]{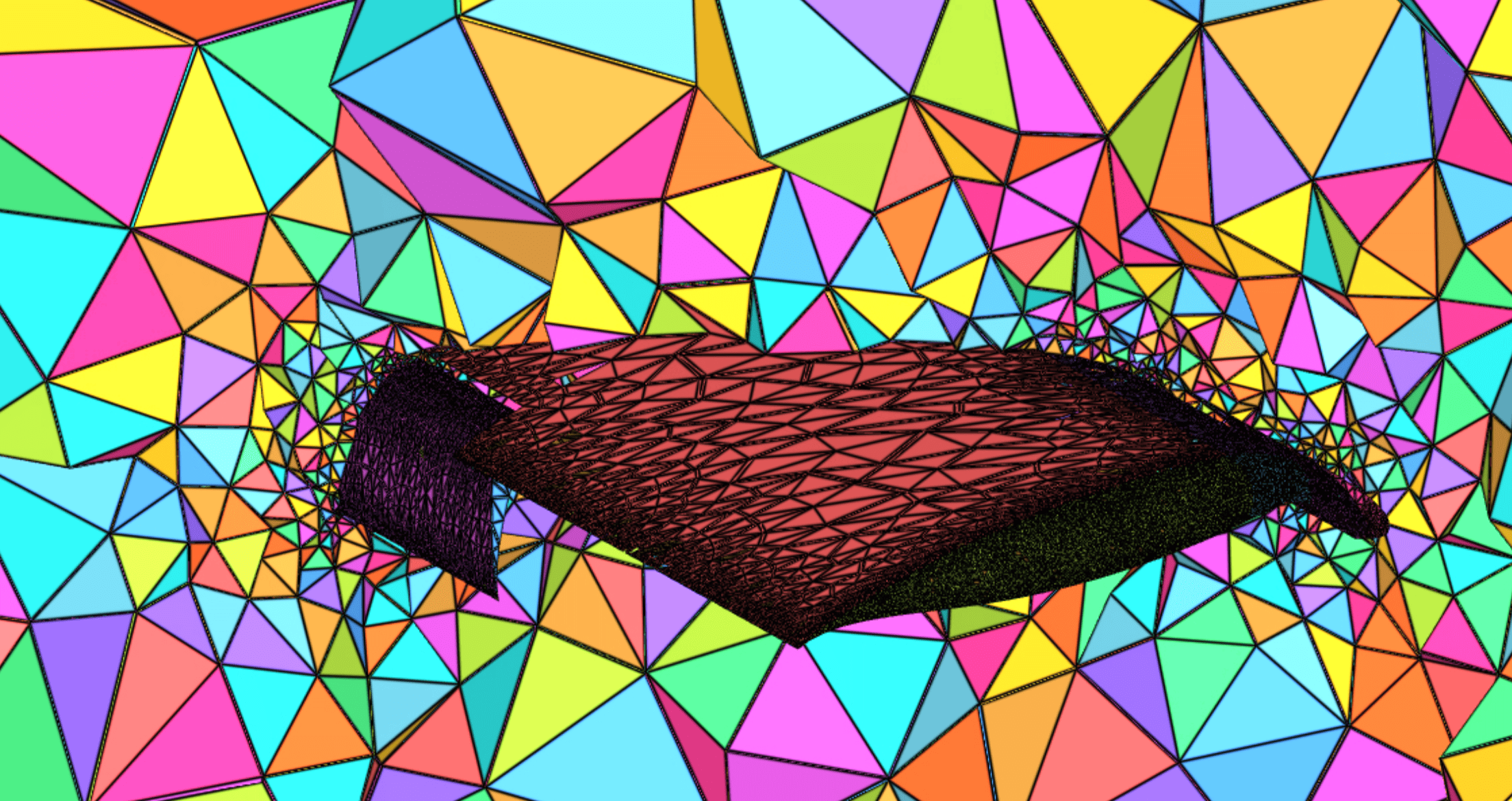}
  \caption{Slices of the resulting pentatopizations for the rotating CRM-HLS at $t = 0$ (left), $t = 0.0184$ (middle) and $t = 0.0368$ (right) seconds.}
  \label{fig:spacetime-crm-hls}
\end{figure*}
\paragraph{Rotating Hockey Puck.} The next case involves a rotating hockey puck (3 inches in diameter and 1 inch thick).
The puck rotates about the $x$-axis with a time-dependent angle of $\theta(t) = 0.1t$.
Again, 10 time slabs were used to discretize the temporal direction of the domain.
This case also results in a pentatopization that fully conforms to the input tetrahedralization.
The initial Delaunay mesh only contains about 32\% of the 23,675 boundary tetrahedral constraints (it misses 16,109 tetrahedra).
Again, the first iteration of the ARO algorithm gets very close to the input boundary mesh, and only misses 120 constraints.
After a total of 5 iterations of combining the ARO algorithm with the addition of Steiner vertices (153 in total), the algorithm terminates with no missing boundary faces.
Slices of the pentatopizations are shown in \cref{fig:spacetime-puck} at the initial, middle and final times of the spacetime domain.
\paragraph{Inflating Pool Toy.} This example represents the time-dependent inflation of the duck-shaped pool toy of Fig.~\ref{fig:surfaces-3d}~\cite{Crane_2013}.
In this example, only 1 time slab is used.
This is partially motivated by the fact that future work may parallelize the boundary recovery across time slabs after generating tetrahedralizations at each time step.
However, this will definitely require suppressing Steiner vertices by either deleting them or displacing them to the interior volume.
The use of a single slab is also motivated by the fact that the Steiner vertex insertion procedure is the slowest part of the overall boundary recovery procedure.
The expansion of the model is very mild, with points scaled at time $t$ according to $(1 + t)$ with a time step of 0.030165 seconds.
The initial Delaunay mesh contains 71\% of the boundary tetrahedra (17,491 tetrahedra are missing).
After the first iteration of the ARO algorithm, only 108 tetrahedra are missing, demonstrating its effectiveness for recovering the boundary tetrahedra.
After a few iterations of adding Steiner vertices and restarting the ARO algorithm, a complete boundary-conforming mesh (with 112 Steiner vertices) is obtained.
Slices of the pentatopizations are shown in \cref{fig:spacetime-duck} at the initial, middle and final times of the spacetime domain.
\paragraph{Expanding Stanford Bunny and Spot.} Despite the success of the ARO algorithm so far, it is worth demonstrating a few cases that were not successful in completely conforming to the boundary.
Again, a single-slab tetrahedral extrusion was used with a time step equal to the smallest edge length in the mesh.
Points at time $t$ were expanded by a factor of $(1 + t)$. 
For the expanding Stanford Bunny and Spot models, \cref{alg:aro} (combined with the addition of Steiner vertices) was able to recover 99.991\% and 99.997\% of the respective boundary meshes, but was not able to make further progress.
The Bunny mesh still missed 11 boundary tetrahedra whereas the Spot mesh missed 2 tetrahedra.
This is perhaps due to the preliminary nature of the intersection-based Steiner vertex insertion procedure.
Future work consists of finding a better, more robust method for inserting Steiner vertices when the frontal procedure stalls.
Nonetheless, the first iteration of the ARO algorithm only missed 148 and 50 tetrahedra for the Stanford Bunny and Spot test cases, respectively, after initially missing 33,579 and 18,545 tetrahedra in the initial unconstrained Delaunay mesh, which again demonstrates the effectiveness of this algorithm for recovering boundary faces.
\paragraph{Rotating Simplified High-Lift Common Research Model.} Finally, let us revisit a three-dimensional version of the rotating flap initially presented in \cref{fig:spacetime-meshing}.
Specifically, the geometry for the first test case of the High-Lift Prediction Workshop was meshed isotropically and then extruded in time with 5 time slabs. 
At each time $t$, the flap was rotated about an axis aligned with its leading edge at an angle of $\theta(t) = 10t$.
Again, the time step used in the extrusion was the smallest edge length in the initial surface mesh.
This test case simply focuses on the ARO algorithm over a single iteration (no Steiner vertices were added).
The boundary mesh contains about 8.5 million tetrahedra, and the Delaunay mesh of the boundary vertices only contained 58\% of these.
The ARO procedure was able to recover 99.255\% of these tetrahedra in a single iteration, meaning 61,547 constraints were still missing.
\Cref{fig:spacetime-crm-hls} shows slices of the pentatopizations at the initial (left) and final time steps (right).
The middle picture shows a slice of the pentatopization at the middle of the temporal domain.
This picture also shows that much of the missing constraints occur in the middle of the spacetime domain since there are several holes on the upper surface of the flap.
Note that the Delaunay meshing (DT) time reported for this test case appears inconsistent with the performance analysis of \cref{sec:performance}, but this could be due to the fact that the point distributions are highly irregular and clustered towards the surfaces instead of uniformly distributed as in \cref{sec:performance}.
Furthermore, points were inserted in a frontal manner using the surface tetrahedralization instead of following a Hilbert curve.
\section{Conclusions and Future Work.}
The Advancing-Ridge cavity Operator (ARO) algorithm was successfully demonstrated on several $3d$ test cases and some $4d$ test cases.
In some cases, Steiner vertices were inserted on the boundary, and the ARO procedure was restarted with the modified boundary.
The Steiner vertex insertion algorithm was not the focus of this paper for a few reasons and will be further explored in the future.
First, it is quite costly to compute all the intersections, especially in $4d$ when there are several candidate edges, triangles and tetrahedra in the mesh that may intersect the constraints.
The second issue is related to the robustness of the intersection-based approach.
George et al.~\cite{George_2003} have demonstrated a robust approach for $3d$ boundary recovery (the current implementation here still fails for some complex $3d$ models), but there were indeed some cases in $4d$ that ultimately still missed some boundary constraints.
Nonetheless, the ARO algorithm was successful in recovering boundary constraints in $3d$ and $4d$.
Furthermore, the underlying cavity operator implementation is very efficient and created pentatopes at a rate of 360,000 - 630,000 pentatopes per second.

In the future, it would be beneficial to improve the $4d$ boundary tetrahedralizations by, for example, adapting to the curvature of the moving geometry in the coupled spacetime domain.
Future work also consists of improving the Steiner vertex insertion algorithm so that a boundary-conforming mesh is always, robustly obtained.
In contrast to typical intersection-based algorithms, when a frontal constraint insertion stalls, the constraining faces that block visibility are known and may provide better information about where a Steiner vertex is needed, whether on the surface or in the volume.
An option to either remove or displace Steiner vertices into the volume will also be explored.
Considering the resulting meshes are meant for a spacetime numerical simulation, vertices in the interior of the domain will also be needed to improve the mesh quality.
Slivers will also need to be removed since the resulting meshes can contain cells with near zero volumes.
\section*{Acknowledgements}%
The implementation of the edge-edge, edge-triangle, edge-tetrahedron, and triangle-triangle intersection calculations for inserting Steiner vertices was done with the assistance of artificial intelligence tools. The authors assume responsibility for all content.
\bibliographystyle{siam}
\bibliography{references}

\begin{thebibliography}{10}

\bibitem{Anderson_2025_PhD}
{\sc J.~T. Anderson}, {\em Constrained hypervolume boundary meshing techniques
  for four-dimensional space-time applications}, {PhD} thesis, Pennsylvania
  State University, May 2025.

\bibitem{Anderson_2023}
{\sc J.~T. Anderson, D.~M. Williams, and A.~Corrigan}, {\em Surface and
  hypersurface meshing techniques for space–time finite element methods},
  Comput. Aided Des., 163 (2023).

\bibitem{Balan_2020}
{\sc A.~Balan, M.~A. Park, W.~K. Anderson, D.~S. Kamenetskiy, J.~A. Krakos,
  T.~Michal, and F.~Alauzet}, {\em Verification of anisotropic mesh adaptation
  for turbulent simulations over {ONERA M6} wing}, AIAA J., 58 (2020),
  pp.~1550--1565.

\bibitem{Behr_2008}
{\sc M.~Behr}, {\em Simplex space–time meshes in finite element simulations},
  Int. J. Numer. Methods Fluids, 57 (2008), pp.~1421--1434.

\bibitem{Caplan_2019}
{\sc P.~C. Caplan}, {\em Four-dimensional anisotropic mesh adaptation for
  spacetime numerical simulations}, {PhD} thesis, Massachusetts Institute of
  Technology, June 2019.

\bibitem{Caplan_2022}
\leavevmode\vrule height 2pt depth -1.6pt width 23pt, {\em Parallel
  four-dimensional anisotropic mesh adaptation}, in Proceedings of the 30th
  International Meshing Roundtable, Feb. 2022.

\bibitem{Caplan_2025}
\leavevmode\vrule height 2pt depth -1.6pt width 23pt, {\em Tessellation and
  interactive visualization of four-dimensional spacetime geometries}, Comput.
  Aided Des., 178 (2025).

\bibitem{Caplan_2020}
{\sc P.~C. Caplan, R.~Haimes, D.~L. Darmofal, and M.~C. Galbraith}, {\em
  Four-dimensional anisotropic mesh adaptation}, Comput. Aided Des., 129
  (2020).

\bibitem{Chazelle_1984}
{\sc B.~Chazelle}, {\em Convex partitions of polyhedra: A lower bound and
  worst-case optimal algorithm}, SIAM J. Comput., 13 (1984), pp.~488--507.

\bibitem{Chen_2017}
{\sc J.~Chen, J.~Zheng, Y.~Zheng, H.~Si, O.~Hassan, and K.~Morgan}, {\em
  Improved boundary constrained tetrahedral mesh generation by shell
  transformation}, Appl. Math. Model., 51 (2017), pp.~764--790.

\bibitem{CohenSteiner_2004}
{\sc D.~Cohen-Steiner, Éric Colin de Verdière, and M.~Yvinec}, {\em
  Conforming {D}elaunay triangulations in {3D}}, Comput. Geom., 28 (2004),
  pp.~217--233.

\bibitem{Coupez_2000b}
{\sc T.~Coupez}, {\em {G\'{e}n\'{e}ration de Maillage et Adaptation de Maillage
  par Optimisation Locale}}, Revue Europ\'{e}enne des \'{E}l\'{e}ments Finis, 9
  (2000), pp.~403--423.

\bibitem{Coupez_2000a}
{\sc T.~Coupez, H.~Digonnet, and R.~Ducloux}, {\em {Parallel Meshing and
  Remeshing}}, Applied Mathematical Modelling, 25 (2000), pp.~153--175.

\bibitem{Crane_2013}
{\sc K.~Crane, U.~Pinkall, and P.~Schr{\"o}der}, {\em Robust fairing via
  conformal curvature flow}, ACM Trans. Graph., 32 (2013), pp.~1--10.

\bibitem{Diazzi_2023}
{\sc L.~Diazzi, D.~Panozzo, A.~Vaxman, and M.~Attene}, {\em Constrained
  {D}elaunay tetrahedrization: A robust and practical approach}, ACM Trans.
  Graph., 42 (2023).

\bibitem{Drakopoulos_2017}
{\sc F.~Drakopoulos}, {\em Finite element modeling driven by health care and
  aerospace applications}, {PhD} thesis, Old Dominion University, Aug. 2017.

\bibitem{Erickson_2005}
{\sc J.~Erickson, D.~Guoy, J.~M. Sullivan, and A.~\"{U}ng\"{o}r}, {\em Building
  spacetime meshes over arbitrary spatial domains}, Eng. Comput., 20 (2005),
  p.~342–353.

\bibitem{George_1991}
{\sc P.~George, F.~Hecht, and E.~Saltel}, {\em Automatic mesh generator with
  specified boundary}, Comput. Methods Appl. Mech. Eng., 92 (1991),
  pp.~269--288.

\bibitem{George_2003}
{\sc P.~L. George, H.~Borouchaki, and E.~Saltel}, {\em ``{U}ltimate''
  robustness in meshing an arbitrary polyhedron}, Int. J. Numer. Methods Eng.,
  58 (2003), pp.~1061--1089.

\bibitem{Haimes_2013_ESP}
{\sc R.~Haimes and J.~Dannenhoffer}, {\em The {E}ngineering {S}ketch {P}ad: A
  solid-modeling, feature-based, web-enabled system for building parametric
  geometry}, in 21st AIAA Computational Fluid Dynamics Conference, 2013.

\bibitem{Hu_2020}
{\sc Y.~Hu, T.~Schneider, B.~Wang, D.~Zorin, and D.~Panozzo}, {\em Fast
  tetrahedral meshing in the wild}, ACM Trans. Graph., 39 (2020).

\bibitem{Hu_2018}
{\sc Y.~Hu, Q.~Zhou, X.~Gao, A.~Jacobson, D.~Zorin, and D.~Panozzo}, {\em
  Tetrahedral meshing in the wild}, ACM Trans. Graph., 37 (2018).

\bibitem{Loseille_2021}
{\sc A.~Loseille}, {\em {Mesh generation and adaptation for scientific
  computing}}, {H}abilitation {\`a} diriger des recherches, {Universit{\'e}
  Paris-Saclay}, March 2021.

\bibitem{Loseille_2017}
{\sc A.~Loseille, F.~Alauzet, and V.~Menier}, {\em Unique cavity-based operator
  and hierarchical domain partitioning for fast parallel generation of
  anisotropic meshes}, Comput. Aided Des., 85 (2017).

\bibitem{Loseille_2018}
{\sc A.~Loseille and R.~Feuillet}, {\em Vizir: High-order mesh and solution
  visualization using {OpenGL} 4.0 graphic pipeline}, in 2018 AIAA Aerospace
  Sciences Meeting.

\bibitem{Levy_2016}
{\sc B.~Lévy}, {\em Robustness and efficiency of geometric programs: The
  {P}redicate {C}onstruction {K}it ({PCK})}, Comput. Aided Des., 72 (2016),
  pp.~3--12.

\bibitem{Lohner_1988}
{\sc R.~Löhner and P.~Parikh}, {\em Generation of three-dimensional
  unstructured grids by the advancing-front method}, Int. J. Numer. Methods
  Fluids, 8 (1988), pp.~1135--1149.

\bibitem{Marcum_2001}
{\sc D.~Marcum}, {\em Efficient generation of high-quality unstructured surface
  and volume grids}, Eng. Comput., 17 (2001), pp.~211--233.

\bibitem{Marcum_2014}
{\sc D.~Marcum and F.~Alauzet}, {\em Aligned metric-based anisotropic solution
  adaptive mesh generation}, Procedia Eng., 82 (2014), pp.~428--444.
\newblock 23rd International Meshing Roundtable.

\bibitem{Marot_2018}
{\sc C.~Marot, J.~Pellerin, and J.-F. Remacle}, {\em One machine, one minute,
  three billion tetrahedra}, Int. J. Numer. Methods Eng., 117 (2019),
  pp.~967--990.

\bibitem{Peraire_1988}
{\sc J.~Peraire, J.~Peiro, L.~Formaggia, K.~Morgan, and O.~C. Zienkiewicz},
  {\em Finite element {E}uler computations in three dimensions}, Int. J. Numer.
  Methods Eng., 26 (1988), pp.~2135--2159.

\bibitem{Pirzadeh_1993}
{\sc S.~Pirzadeh}, {\em Unstructured viscous grid generation by advancing-front
  method}, NASA Contractor Report 191449, NASA, 1993.

\bibitem{Schonhardt_1928}
{\sc E.~Sch{\"o}nhardt}, {\em {\"U}ber die zerlegung von dreieckspolyedern in
  tetraeder}, Math. Ann., 98 (1928), pp.~309--312.

\bibitem{Shewchuk_1996_Triangle}
{\sc J.~R. Shewchuk}, {\em Triangle: {E}ngineering a {2D} quality mesh
  generator and {D}elaunay triangulator}, in Applied Computational Geometry
  Towards Geometric Engineering, 1996, pp.~203--222.

\bibitem{Shewchuk_2000}
\leavevmode\vrule height 2pt depth -1.6pt width 23pt, {\em Sweep algorithms for
  constructing higher-dimensional constrained {D}elaunay triangulations}, in
  Proceedings of the Sixteenth Annual Symposium on Computational Geometry,
  2000, p.~350–359.

\bibitem{Shewchuk_2008}
\leavevmode\vrule height 2pt depth -1.6pt width 23pt, {\em General-dimensional
  constrained {D}elaunay and constrained regular triangulations, {I}:
  {C}ombinatorial properties}, Discrete Comput. Geom., 39 (2008), p.~580–637.

\bibitem{Si_2015}
{\sc H.~Si}, {\em {TetGen}, a {D}elaunay-based quality tetrahedral mesh
  generator}, ACM Trans. Math. Softw., 41 (2015).

\bibitem{Si_2011}
{\sc H.~Si and K.~Gärtner}, {\em {3D} boundary recovery by constrained
  {D}elaunay tetrahedralization}, Int. J. Numer. Methods Eng., 85 (2011),
  pp.~1341--1364.

\bibitem{Si_2013}
{\sc H.~Si and J.~R. Shewchuk}, {\em Incrementally constructing and updating
  constrained {D}elaunay tetrahedralizations with finite precision
  coordinates}, in Proceedings of the 21st International Meshing Roundtable,
  2013, pp.~173--190.

\bibitem{Turk_1994}
{\sc G.~Turk and M.~Levoy}, {\em Zippered polygon meshes from range images}, in
  Proceedings of the 21st Annual Conference on Computer Graphics and
  Interactive Techniques, 1994, pp.~311--318.

\bibitem{Ungor_2000}
{\sc A.~{\"U}ng{\"o}r and A.~Sheffer}, {\em {Tent-Pitcher}: {A} meshing
  algorithm for space-time discontinuous {G}alerkin methods}, in Proceedings of
  the 9th International Meshing Roundtable, 2000, pp.~111--122.

\bibitem{vonDanwitz_2021}
{\sc M.~von Danwitz, P.~Antony, F.~Key, N.~Hosters, and M.~Behr}, {\em
  Four-dimensional elastically deformed simplex space-time meshes for domains
  with time-variant topology}, Int. J. Numer. Methods Fluids, 93 (2021),
  pp.~3490--3506.

\bibitem{Watson_1981}
{\sc D.~F. Watson}, {\em Computing the $n$-dimensional {D}elaunay tessellation
  with application to {V}oronoi polytopes}, Comput. J., 24 (1981), p.~167.

\bibitem{Weatherill_1994}
{\sc N.~P. Weatherill and O.~Hassan}, {\em Efficient three-dimensional
  {D}elaunay triangulation with automatic point creation and imposed boundary
  constraints}, Int. J. Numer. Methods Eng., 37 (1994), pp.~2005--2039.

\end{thebibliography}
\end{document}